\documentclass[12pt,english]{article}
\usepackage[T1]{fontenc}
\usepackage[latin9]{inputenc}
\usepackage{float}
\usepackage{mathrsfs}
\usepackage{amsmath}
\usepackage{amssymb}
\usepackage{graphicx}
\usepackage{esint}

\makeatletter

\date{}

\newcommand{\Pibar}{\overline{\Pi}}
\newcommand{\abar}{\bar{a}}
\newcommand{\bbar}{\bar{b}}
\newcommand{\xibar}{\bar{\xi}}
\newcommand{\taubar}{\bar{\tau}}
\newcommand{\Omegabar}{\bar{\Omega}}
\newcommand{\Xbar}{\overline{X}}
\newcommand{\Ybar}{\overline{Y}}
\newcommand{\Zbar}{\overline{Z}}

\newcommand{\Wbar}{\overline{W}}
\newcommand{\Fbar}{\overline{F}}

\usepackage{babel}

\usepackage{babel}

\usepackage{babel}

\@ifundefined{showcaptionsetup}{}{%
 \PassOptionsToPackage{caption=false}{subfig}}
\usepackage{subfig}

\usepackage{babel}

\makeatother

\usepackage{babel}
\begin{document}

\title{Warped Vacuum Statistics}

\author{Pontus Ahlqvist%
\thanks{pontus@phys.columbia.edu%
}, Brian R. Greene%
\thanks{greene@phys.columbia.edu%
}, and David Kagan%
\thanks{dk296@columbia.edu%
}\\
 \\
 \emph{Institute of Strings, Cosmology, and Astroparticle Physics}\\
 \emph{Department of Physics}\\
 \emph{Columbia University, New York, NY 10027, USA}}
\maketitle
\begin{abstract}
We consider the effect of warping on the distribution of type IIB
flux vacua constructed with Calabi-Yau orientifolds. We derive an
analytical form of the distribution that incorporates warping and
find close agreement with the results of a Monte Carlo enumeration
of vacua. Compared with calculations that neglect warping, we find
that for any finite volume compactification, the density of vacua
is highly diluted in close proximity to the conifold point, with a
steep drop-off within a critical distance. 
\end{abstract}
\global\long\def\Omegabar{\bar{\Omega}}

\global\long\def\Xbar{\overline{X}}

\global\long\def\Ybar{\overline{Y}}

\global\long\def\Zbar{\overline{Z}}

\global\long\def\xibar{\bar{\xi}}
 \global\long\def\taubar{\overline{\tau}}
 \global\long\def\Wbar{\overline{W}}
 \global\long\def\abar{\overline{a}}
 \global\long\def\bbar{\overline{b}}
 \global\long\def\Fbar{\overline{F}}
 \global\long\def\Pibar{\overline{\Pi}}

\section{Introduction}

Complex structure moduli in type IIB string theory are stabilized
by turning on fluxes, and in certain parts of the theory's moduli
space the fluxes lead to large warping effects. These effects are
essential for a detailed understanding of dynamics in the string theory
landscape \cite{Conifunneling}. Tunneling between flux vacua involves
the nucleation of a brane carrying appropriate charges, but such events
appear to be favored in configurations of the Calabi-Yau geometry
where particular cycles are small, and hence, warping due to fluxes
through such cycles is large.

Aside from dynamics, the distribution of vacua for such models is
of interest. The Bousso-Polchinski model of the string landscape \cite{BoussoPolchinski}
suggests that with a sufficient number of fluxes, one should expect
vacuum energies that are sufficiently finely spaced to ensure that
some vacua have cosmological constants in rough agreement with our
own. In \cite{DouglasAshok,DouglasDenef}, the authors developed a
convenient framework for carrying out such analyses in the context
of type IIB string theory compactifications. The theoretical vacuum
distributions for certain simple Calabi-Yau compactifications have
also been supported by numerical studies \cite{KachruTaxonomy}.

A quite general result of these studies is that vacua appear to accumulate
around the conifold point in the complex structure moduli space. In
fact, the density of these vacua diverges logarithmically. However,
in light of the fact that warping becomes strong precisely near the
conifold point for any finite volume Calabi-Yau compactification,
the natural question arises of what effect---if any---warping may
have on the distribution of vacua.

A related point is whether the vacuum density is well-captured by
the simpler to compute index density. In the absence of warping, the
index is simpler to compute since it is similar to the Chern class
of the moduli space, whereas there is no straightforward geometric
quantity that corresponds to the vacuum count. As we shall see, when
warping is included the overall agreement between number and index
densities continues to hold, but the topological nature of the latter
becomes more complicated.

In section \ref{sec:Analytical-Distributions} we review the general
framework for deriving theoretical distributions of vacua for unwarped
Calabi-Yau compactifications as originally laid out in \cite{DouglasDenef}.
We then explain how to modify this construction to derive the warped
version of the number and index densities. The methods are then used
to explicitly compute the densities in the vicinity of a conifold
point. In section \ref{sec:Numerical-Vacuum-Statistics} we numerically
generate near-conifold distributions of vacua and compare these to
the theoretical distributions of section \ref{sec:Analytical-Distributions}.

\section{Background}

Type IIB string theory compactified on a Calabi-Yau manifold yields
scalar fields, known as moduli, in the low energy supergravity theory.
These moduli are related to geometrical parameters of the internal
Calabi-Yau manifold and can in certain models number in the hundreds.
Unfortunately, these moduli appear as massless fields without any
potential governing their dynamics, rendering the physics unrealistic.
Fortunately string theory contains other ingredients with the capacity
to resolve this problem. In particular, these scalar fields can be
stabilized by turning on various p-form fluxes in the internal manifold,
a procedure that generates the Gukov-Vafa-Witten superpotential:
\begin{equation}
W(z)=\int\Omega_{3}\wedge G_{3}\label{GVW-Potential}
\end{equation}
Here, $\Omega_{3}$ is the holomorphic $(3,0)$ form defined on the
Calabi-Yau, $G_{3}=F_{3}-\tau H_{3}$ is the type IIB $3$-form field
strength, $\tau$ is the axio-dilaton, and $z$ denotes the set of
complex moduli mentioned above upon which the holomorphic three form
depends. Given this superpotential, the scalar potential for the moduli
becomes:
\begin{equation}
V(z,\tau)=e^{K/M_{P}^{2}}\left(K^{a\overline{b}}D_{a}W\,\overline{D_{b}W}-\frac{3}{M_{P}^{2}}|W|^{2}\right)\label{Scalar-Potential}
\end{equation}
where the sum runs over the complex moduli ($i,j=1,2,\ldots,n$),
with $n=h_{CY}^{2,1}$, as well as the axio-dilaton ($i,j=0$). Here,
the covariant derivative acts as $D_{a}=\partial_{a}+K_{a}$ where
$K_{a}$ is the derivative of the Kähler potential with repsect to
the $a^{\textrm{th}}$ complex modulus or $\tau$. This ensures that
$DW$ transforms in the same way as $W$ itself under a Kähler transformation
so that the physical potential $V(z,\tau)$ is invariant under Kähler
transformations. Supersymmetric minima of the potential $V$ occur
at points in moduli space where $D_{a}W=0$ with $a$ running over
all of the moduli and the axio-dilaton. In general, the potential
has many minima, each of which represents a stable low energy configuration
of the internal Calabi-Yau. These configurations arise from the large
number of discrete fluxes that can thread through the Calabi-Yau's
various 3-cycles. It is thus natural to explore this large landscape
of flux vacua using statistical methods, as was first done in \cite{DouglasAshok,DouglasDenef},
as we now briefly review.

\section{Analytical Distributions\label{sec:Analytical-Distributions}}

\subsection{Counting the vacua\label{sub:Counting-the-vacua}}

Here we will review the derivation of the index density given by Douglas
and Denef in \cite{DouglasDenef}, focusing on areas where our analysis,
including the effects of warping, will differ. We will restrict attention
to vacua that satisfy $D_{a}W=0$ for all complex moduli and the axio-dilaton.
The strategy is to consider these equations as constraints on the
choice of fluxes and otherwise, simply allow the fluxes to scan. First,
assume that fluxes are fixed and consider the function on moduli space
given by%
\footnote{Our conventions for the delta functions and integration measures depending
on a complex variable $z$ are given by $\delta^{2}(z)=\delta(\textrm{Re }z)\delta(\textrm{Im }z)$,
and $d^{2}z=d(\textrm{Re }z)d(\textrm{Im }z)$.%
}
\begin{equation}
\delta^{2n+2}(DW(z))\equiv\delta(D_{0}W(z))\ldots\delta(D_{n}W(z))\delta(\overline{D_{0}W(z)})\ldots\delta(\overline{D_{n}W(z)}).\label{deltas}
\end{equation}
Clearly this provides support only at the locations of the vacua.
However, as written each vacuum does not contribute with the same
weight. To see this, rewrite equation (\ref{deltas}) as a sum of
delta functions which explicitly spike at the locations of the minima:
\begin{equation}
\delta^{2n+2}(DW(z))=\sum_{\text{vac}}\frac{\delta^{2n+2}(z-z_{\text{vac}})}{|\det D^{2}W|}.\label{delta-rewrite}
\end{equation}
Here the determinant arises from expanding the delta functions near
each minimum in much the same way as $\delta(f(x))=\sum\delta(x-x_{\text{zero}})/|f'(x)|$,
and is of the $(2n+2)\times(2n+2)$ matrix
\begin{equation}
\begin{pmatrix}\partial_{a}D_{b}W &  & \partial_{a}\overline{D_{b}W}\\
\overline{\partial}_{\overline{a}}D_{b}W &  & \overline{\partial_{a}D_{b}W}
\end{pmatrix},\label{fermMassMatrix}
\end{equation}
where we let $a,b$ range over the $n$ moduli as well as the axio-dilaton.
Note that the partial deriviatives in the matrix above can be replaced
by covariant derivatives at the vacua since there the conditions $D_{a}W=0$
render the two expressions equivalent. If we then integrate this over
the moduli space we find contributions from each vacuum associated
with a fixed set of fluxes with weight $|\det D^{2}W|^{-1}$. Since
this value is not constant over the moduli space, the result will
not reflect the number of vacua. To count the vacua, we must compensate
by integrating over the delta-functions appropriately weighted:
\begin{equation}
\int d^{2n}zd^{2}\tau\hspace{1mm}\delta^{2n+2}(DW(z))|\det D^{2}W|.\label{count}
\end{equation}
This expression defines the vacuum count for a given set of fluxes.
Another useful quantity considered in \cite{DouglasDenef} is the
index, which involves dropping the absolute values around the determinant
of the fermion mass matrix:
\begin{equation}
\int d^{2n}zd^{2}\tau\hspace{1mm}\delta^{2n+2}(DW(z))\det D^{2}W.\label{index}
\end{equation}
This integral then counts the number of positive vacua minus the number
of negative vacua, where parity is given by the sign of the determinant
of the matrix in equation (\ref{fermMassMatrix}). To count all vacua,
we must also sum over fluxes subject to the tadpole cancellation condition
\begin{equation}
L=\int_{\text{CY}}F_{3}\wedge H_{3}\leq L_{*}.\label{tadpole}
\end{equation}
Here $L_{*}$ is the maximum possible value for $L$. It will turn
out to be useful to lift this discussion to F-theory where we consider
our manifold $\mathcal{M}$ as an elliptically fibered Calabi-Yau
4-fold, whose base consists of the original 3-fold and fibers are
given by the auxiliary 2-torus whose period is given by the axio-dilaton
$\tau$. We decompose the holomorphic 4-form:
\begin{equation}
\Omega_{4}=\Omega_{1}\wedge\Omega_{3},
\end{equation}
 where $\Omega_{1}$ is the holomorphic one form on the two torus
parameterizing the axio-dilaton, and $\Omega_{3}$ is the usual holomorphic
three form on the Calabi-Yau. In particular, if we consider the two
one-cylces $\mathcal{A}$ and $\mathcal{B}$ on the torus, we can
define the two one forms $\alpha$ and $\beta$ dual to the cycles
$\mathcal{A}$ and $\mathcal{B}$ such that $\int_{\mathcal{A}}\gamma=\int_{T^{2}}\alpha\wedge\gamma$
and $\int_{\mathcal{B}}\gamma=\int_{T^{2}}\beta\wedge\gamma$ for
all closed one forms $\gamma$. Then, as long as we define our holomorphic
one-form $\Omega_{1}$ as
\begin{equation}
\Omega_{1}=\alpha-\tau\beta,
\end{equation}
we will have $\tau=\int_{\mathcal{A}}\Omega_{1}/\int_{\mathcal{B}}\Omega_{1}$
as we want for the complex structure of the torus. Furthermore, if
we define a flux four form as $G_{4}=\beta\wedge F_{3}-\alpha\wedge H_{3}$,
we can write the tadpole condition as
\begin{equation}
\frac{1}{2}\int_{\mathcal{M}}G_{4}\wedge G_{4}=-\int_{T^{2}}\alpha\wedge\beta\int_{CY_{3}}F_{3}\wedge H_{3}
\end{equation}

If we normalize the F-theory torus volume so that $\int_{T^{2}}\alpha\wedge\beta=-1$,
this exactly reproduces the tadpole condition in the type IIB picture.
With $K=\dim H_{CY_{3}}^{3}$ we've lumped the $2K$ fluxes $F_{0},\ldots,F_{K-1},H_{0},\ldots,H_{K-1}$
into the $2K$ components of $G_{4}$. Also note that with this definition
of the flux four form, we can write the usual type IIB superpotential
as
\begin{equation}
W=\int_{\mathcal{M}}\Omega_{4}\wedge G_{4}.
\end{equation}

Let's choose a particular basis of three forms on the $\textrm{CY}_{3}$
$\{\Sigma_{i}\}$, and denote the intersection form in this basis
as $Q_{ij}$ so that
\begin{equation}
\int_{CY_{3}}\Sigma_{i}\wedge\Sigma_{j}=Q_{ij}
\end{equation}
We can extend this basis to $\mathcal{M}$ by wedging it with the
one forms $\alpha$ and $\beta$. In this basis, we denote the components
of the field strength $G_{4}$ by $N_{a}$ with $a=0,1,\ldots2K-1$,
and the intersection form in the full 4 (complex) dimensional space
by $\eta_{ab}$. Then, the tadpole condition in equation (\ref{tadpole})
can be written in terms of the components of the two fluxes ($F=F^{i}\Sigma_{i}$
and $H=H^{i}\Sigma_{i}$) as
\begin{equation}
L=\frac{1}{2}N^{a}\eta_{ab}N^{b}=F^{i}Q_{ij}H^{j}\leq L_{*}
\end{equation}
We should then sum only over the fluxes that satisfy this inequality.
In particular we can imagine summing over all fluxes while including
a step function.
\begin{equation}
\text{Index}=\sum_{\text{Fluxes}}\theta(L_{*}-L)\int d^{2n}zd^{2}\tau\hspace{1mm}\delta^{2n+2}(DW(z))\det D^{2}W
\end{equation}
We can write the step function as an integral over a delta function%
\footnote{Note that in \cite{DouglasDenef} the step-function is expressed in
terms of a contour integral over an exponential $e^{\alpha L_{*}}$.
Our expression in terms of a delta function proves to be more useful
for the analysis incorporating warping effects.%
},
\begin{equation}
\theta(L_{*}-L)=\int_{-\infty}^{L_{*}}\delta(L-\widetilde{L})d\widetilde{L}
\end{equation}
yielding
\begin{equation}
\text{Index}=\sum_{\text{Fluxes}}\int_{-\infty}^{L_{*}}d\widetilde{L}\int d^{2n}zd^{2}\tau\hspace{1mm}\delta(L-\widetilde{L})\delta^{2n+2}(DW(z))\det D^{2}W
\end{equation}
By treating the fluxes $N_{0},\ldots,N_{2K-1}$ as continuously varying
parameters, we can approximate this sum by an integral,
\begin{equation}
\text{Index}=\int_{-\infty}^{L_{*}}d\widetilde{L}\int d^{2K}N\int d^{2n}zd^{2}\tau\hspace{1mm}\delta(L-\widetilde{L})\delta^{2n+2}(DW(z))\det D^{2}W
\end{equation}

It is natural to define the \emph{index density} in moduli (and axio-dilaton)
space by
\begin{equation}
\mu_{I}(z,\tau)=\int_{-\infty}^{L_{*}}d\widetilde{L}\int d^{2K}N\hspace{1mm}\delta(L-\widetilde{L})\delta^{2n+2}(DW(z))\det D^{2}W
\end{equation}

Upon integrating over $\tau,z$, this will then equal the total index.
We now rewrite this index density in terms of geometric properties
of the moduli space. A first step in doing this is to change basis
from $\left\{ \alpha\wedge\Sigma_{a},\beta\wedge\Sigma_{a}\right\} $
to the set of linearly independent four forms $\{\Omega_{4},D_{a}\Omega_{4},D_{0}D_{i}\Omega_{4}\}\cup\{c.c\}$
where $a$ ranges over the complex moduli as well as the axio-dilaton
while $i$ ranges only over the moduli. This proposed basis consists
of $4(n+1)$ elements where $n$ denotes the number of complex moduli
in our theory, which agrees with the 2$K$ elements of the original
basis. This new basis satisfies
\begin{eqnarray}
\int_{\mathcal{M}}\Omega_{4}\wedge\bar{\Omega}_{4} & = & e^{-K(\tau,z)}\\
\int_{\mathcal{M}}D_{a}\Omega_{4}\wedge\bar{D}_{\bar{b}}\bar{\Omega}_{4} & = & -e^{-K(\tau,z)}K_{a\bar{b}}\\
\int_{\mathcal{M}}D_{0}D_{i}\Omega_{4}\wedge\bar{D}_{\bar{0}}\bar{D}_{\bar{j}}\bar{\Omega}_{4} & = & e^{-K(\tau,z)}K_{\tau\bar{\tau}}K_{i\bar{j}},
\end{eqnarray}
with all other combinations vanishing. By rescaling all of our basis
elements by the factor $e^{K(\tau,z)/2}$, the new basis won't have
any of the extra exponentials in their inner products:
\begin{eqnarray}
\int_{\mathcal{M}}e^{K(\tau,z)/2}\Omega_{4}\wedge e^{K(\tau,z)/2}\bar{\Omega}_{4} & = & 1\\
\int_{\mathcal{M}}e^{K(\tau,z)/2}D_{i}\Omega_{4}\wedge e^{K(\tau,z)/2}\bar{D}_{\bar{j}}\bar{\Omega}_{4} & = & -K_{i\bar{j}}\\
\int_{\mathcal{M}}e^{K(\tau,z)/2}D_{0}D_{i}\Omega_{4}\wedge e^{K(\tau,z)/2}\bar{D}_{\bar{0}}\bar{D}_{\bar{j}}\bar{\Omega}_{4} & = & K_{\tau\bar{\tau}}K_{i\bar{j}},
\end{eqnarray}
And becuase of the properties of the covariant derivative, we can
accomplish these changes by rescaling the holomorphic 4-form by this
same factor: $\Omega_{4}\rightarrow e^{K(\tau,z)/2}\Omega_{4}$. For
notational simplicity we will redefine $\Omega_{4}$ to represent
this rescaled version%
\footnote{ The covariant derivative $D_{a}=\partial_{a}+K_{a}$, is the Hermitian
metric connection acting on sections of the complex line bundle $L$,
where $\Omega_{3}$ is a section of $H\otimes L$, with $H$ the Hodge
bundle and $L$ is a line bundle whose first Chern class is the Kähler
form on the 3-fold's moduli space. The expression $\int\Omega_{3}\wedge\Omegabar_{3}$
provides a metric on $L$ from which the metric connection then follows.
When acting on sections of other, related bundles, the Hermitian metric
connection must be appropriately modified.%
}.When we want to explicitly refer to the actual holomorphic 4-form,
we will denote it as $\widehat{\Omega}_{4}$:
\begin{equation}
\Omega_{4}=e^{K(\tau,z)/2}\widehat{\Omega}_{4}
\end{equation}

Finally, we can consider the set $\mathcal{B}=\{\Omega_{4},D_{A}\Omega_{4},D_{\underline{0}}D_{I}\Omega_{4}\}\cup\{c.c.\}$
where $D_{A}\equiv e_{A}^{a}D_{a}$, and the vielbeins $e_{A}^{a}$
satisfy $e_{A}^{a}e_{\bar{B}}^{\bar{b}}K_{a\bar{b}}=\delta_{A\bar{B}}$,
as usual. The notation is consistent assuming a suitably defined spin-connection
(see appendix \ref{sub:Covariant-Derivatives}). Our new basis is
orthonormal:
\begin{eqnarray}
\int_{\mathcal{M}}\Omega_{4}\wedge\bar{\Omega}_{4} & = & 1\\
\int_{\mathcal{M}}D_{A}\Omega_{4}\wedge\bar{D}_{\bar{B}}\bar{\Omega}_{4} & = & -\delta_{A\bar{B}}\\
\int_{\mathcal{M}}D_{\underline{0}}D_{I}\Omega_{4}\wedge\bar{D}_{\underline{\bar{0}}}\bar{D}_{\bar{J}}\bar{\Omega}_{4} & = & \delta_{I\bar{J}},
\end{eqnarray}

The $4$-form flux $G_{4}$ in the new basis is given by
\begin{equation}
G_{4}=\Xbar\Omega_{4}-\Ybar{}^{A}D_{A}\Omega_{4}+\Zbar{}^{I}D_{\underline{0}}D_{I}\Omega_{4}+\text{c.c.}\label{eq:G4OmegaBasis}
\end{equation}
with $X,Y^{\bar{A}},Z^{\bar{I}},\Xbar,\Ybar^{A},\Zbar^{I}$ being
the coefficients of $G_{4}$ in this basis. Note that $G_{4}$ does
not depend on the complex structure or axio-dilaton, which implies
that the coefficients $X,Y^{\overline{A}},Z^{\overline{I}},\ldots$
depend on $z^{i}$ and $\tau$ in a way that precisely cancels the
dependences arising from $\Omega_{4}$ and its derivatives. Since
$G_{4}$ doesn't depend on the complex structure of the Calabi-Yau
or the axio-dilaton, we can relate these coefficients to various combinations
of derivatives acting on the superpotential. In particular
\begin{eqnarray}
W & = & \int\Omega_{4}\wedge G_{4}=X\label{eq:Xdef}\\
D_{A}W & = & \int D_{A}\Omega_{4}\wedge G_{4}=Y_{A}\label{eq:Ydef}\\
D_{\underline{0}}D_{\underline{0}}W & = & 0\label{eq:D0D0W0}\\
D_{\underline{0}}D_{I}W & = & \int D_{\underline{0}}D_{I}\Omega_{4}\wedge G_{4}=Z_{I}\label{eq:D0DI}\\
D_{I}D_{J}W & = & \int D_{I}D_{J}\Omega_{4}\wedge G_{4}=\mathcal{F}_{IJK}\Zbar{}^{K}\label{eq:DIDJ}\\
\overline{D}_{\overline{I}}D_{J}W & = & \delta_{\overline{I}J}X\label{eq:DIBDJ}\\
\overline{D}_{\bar{0}}D_{\underline{0}}W & = & X\label{eq:D0bD0}\\
\overline{D}_{\bar{\underline{0}}}D_{I}W & = & 0\label{eq:D0bDI}
\end{eqnarray}
where the computations establishing these relations are provided in
the appendix. Note that we have defined the coefficients $\mathcal{F}_{IJK}=i\int_{CY}\Omega_{3}\wedge D_{I}D_{J}D_{K}\Omega_{3}=i\int_{CY}\Omega_{3}\wedge\partial_{I}\partial_{J}\partial_{K}\Omega_{3}$.
Also, note that $W$ denotes the rescaled superpotential; when we
want to explicitly refer to the original one, we will once again place
a hat on top of it ($\widehat{W}$). We can then rewrite our expressions
in terms of these new functions on moduli space, and in particular
have for the tadpole condition
\begin{equation}
L=\frac{1}{2}N\eta N=\frac{1}{2}\int G_{4}\wedge G_{4}=|X|^{2}-|Y|^{2}+|Z|^{2},
\end{equation}

where $|Y|^{2}=\Ybar^{A}Y^{\bar{A}}\delta_{\bar{A}A}$, etc. The index
density then becomes
\begin{equation}
\begin{aligned}\mu_{I}(z,\tau)=\int_{-\infty}^{L_{*}}d\widetilde{L}\int d^{2}X\, d^{2n+2}Y\, d^{2n}Z\, J\,|\det g|\,\delta(\widetilde{L}-|X|^{2}+|Y|^{2}-|Z|^{2})\delta^{2n+2}(Y_{A})\,|X|^{2}\\
\times\det\begin{pmatrix}\Xbar\delta_{IJ}-\frac{Z_{I}\Zbar_{J}}{X} &  & \mathcal{F}_{IJK}\Zbar{}^{K}\\
\mathcal{\overline{F}}_{IJK}Z^{K} &  & X\delta_{IJ}-\frac{\Zbar_{I}Z_{J}}{\Xbar}
\end{pmatrix}
\end{aligned}
\end{equation}

Here $J$ is the Jacobian obtained in changing variables from $N_{a}$
to $X,Y_{A},Z_{I}$, which we will determine explicitly below. We
have included an additional factor of $|\det g|$ which comes from
transforming both the delta functions and the determinant to the new
variables, and note that factors of $e^{K}$ cancel between the delta
functions and the determinant.

Let's now compute the Jacobian $|J|$. In the original basis, the
components of $G_{4}$ were given by $N_{a}$. We can now write the
$N_{a}$ in the new basis
\begin{equation}
N=\eta^{-1}\left(\Xbar\Pi-\Ybar{}^{A}D_{A}\Pi+\Zbar{}^{I}D_{\underline{0}}D_{I}\Pi+\textrm{c.c.}\right)
\end{equation}
Here the $\Pi$s are the periods of the rescaled holomorphic four
form and are related to the usual ones by a factor of $e^{K/2}$.We
can see from this expression that the change of basis is achieved
by the application of the matrix $M=\eta^{-1}(\Pi,-D_{A}\Pi,D_{0}D_{I}\Pi,c.c.)$.
If we use the convention that $d^{2}z=\frac{1}{2i}dz\wedge d\bar{z}$,
we find that the appropriate Jacobian is
\begin{equation}
J=2^{2(n+1)}|\det M|=4^{n+1}|\det\eta|^{-1/2}|\det M^{\dagger}\eta M|^{1/2}
\end{equation}
We have $M^{\dagger}\eta M=\text{diag}(1,-\mathbf{1}_{n+1},\mathbf{1}_{n},1,-\mathbf{1}_{n+1},\mathbf{1}_{n})$,
which follows from our choice of an orthonormal basis of 4-forms.
This implies that the Jacobian is given by
\begin{equation}
J=4^{n+1}|\det\eta|^{-1/2}.
\end{equation}

The final expression is then (after explicitly integrating over $Y_{A}$),
\begin{align}
\mu_{I}(z,\tau) & =4^{n+1}|\det\eta|^{-1/2}\int_{-\infty}^{L_{*}}d\tilde{L}\int d^{2}X\, d^{2n}Z\,|\det g|\,\delta(\tilde{L}-|X|^{2}-|Z|^{2})|X|^{2}\nonumber \\
 & \times\det\begin{pmatrix}\Xbar\delta_{IJ}-\frac{Z_{I}\Zbar_{J}}{X} &  & \mathcal{F}_{IJK}\Zbar^{K}\\
\mathcal{\overline{F}}_{IJK}Z^{K} &  & X\delta_{IJ}-\frac{\Zbar_{I}Z_{J}}{\Xbar}
\end{pmatrix}.
\end{align}

We can explicitly integrate over the phases, leaving only integrals
over the magnitudes $|X|$ and $|Z|$, showing that the tadpole delta
function fixes the region of integration to lie on a circle of radius
$\sqrt{\widetilde{L}}$ in the $|X|,|Z|$ plane. There is therefore
no need to integrate over negative $\widetilde{L}$s, and furthermore
the remaining finite integral can be evaluated. Following this approach,
one can show that the index density has a nice geometrical interpretation
\cite{DouglasDenef}:
\begin{equation}
\mu_{I}(z,\tau)=\det(R+\omega\mathbb{I}),\label{IndexDensity}
\end{equation}
where $R$ is the curvature two form on the moduli space and $\omega$
is the Kähler form. For the case of one complex modulus (and the axio-dilaton),
this reduces to $\mu_{I}=-\pi^{2}|\det\eta|^{-1/2}\omega_{0}\wedge R_{1}$
where $\omega_{0}$ is the Kähler form on the axio-dilaton side while
$R_{1}$ is the curvature form on the moduli space side. In order
to obtain this, one must use a relationship between the Kähler and
curvature forms on the axio-dilaton moduli space: $R_{0}=-2\omega_{0}$.

\subsection{Incorporating warping\label{sub:Incorporating-warping}}

A full treatment of warped Calabi-Yau geometry involves using the
machinery of generalized complex geometry \cite{Koerber,Martucci,Hitchin}.
However, a rough method that produces the appropriate functional behavior
induced by warping near the conifold will suffice for our purposes.
This behavior can be derived by taking the warped Kähler potential
to be approximated by \cite{Giddings}
\begin{equation}
e^{-\widetilde{K}}=\int e^{-4A}\,\Omega\wedge\Omegabar\approx\int_{\textrm{Bulk}}\Omega\wedge\Omegabar+\int_{\textrm{Conifold}}\left(1+\frac{e^{-4A_{0}}}{c}\right)\Omega\wedge\Omegabar,
\end{equation}
where $e^{-4A}=1+e^{-4A_{0}}/c$ is the warp factor, with $e^{-4A_{0}}$
capturing the significant warping at the conifold while $c$ is a
constant related to the overall volume of the Calabi-Yau manifold.
In general, we will use tildes to denote quantities that include warp
corrections.

The warp-corrected Kähler metric has been shown to have the near-conifold
form \cite{Conifunneling,DouglasWarpedSUSY,DouglasTorroba,DouglasShiu}
\begin{equation}
\widetilde{K}_{\xi\xibar}\approx\frac{K_{1}}{k}-\frac{1}{2\pi k}\log\xi+\frac{C_{w}}{k|\xi|^{4/3}}=K_{\xi\xibar}+\widehat{K}_{\xi\xibar},\label{NearConifoldKahlerMetric}
\end{equation}
where $\xi$ is the local coordinate around the conifold point, $k=\textrm{lim}_{\xi,\xibar\rightarrow0}\, e^{K(\xi,\xibar)}$
and $K_{1}$ is a constant (to leading order) associated with the
Kähler metric's expansion around the conifold. The hatted quantity
in the rightmost expression corresponds to the warp correction to
the original, unwarped Kähler metric. The constant $C_{w}$ is on
the order of the inverse volume of the Calabi-Yau%
\footnote{In fact, it goes approximately like $V_{CY_{3}}^{-2/3}$, since it
is related to the universal Kähler modulus zero mode.%
}, capturing the suppression of the warping effects at large volume.

Given the form of the Kähler metric near the conifold (\ref{NearConifoldKahlerMetric}),
we find that up to shifts by functions holomorphic and antiholomorphic
in $\xi$, we have
\begin{eqnarray}
\widetilde{K} & \approx & K+9C_{w}|\xi|^{2/3}=K+\widehat{K},\label{WarpedKahler}\\
\widetilde{K}_{\xi} & \approx & K_{\xi}+3C_{w}\frac{\xibar^{1/3}}{\xi^{2/3}}=K_{\xi}+\widehat{K}_{\xi}.\label{WarpedKahlerDerivative}
\end{eqnarray}

To take warping into account in computing the vacuum count and index,
we follow the basic logic of section \ref{sub:Counting-the-vacua}
with incorporating various necessary modifications. First, we continue
to define quantities such as $X,Y,$ and $Z$ without making any reference
to the warping. This means that the logic for converting the step
function $\theta(L_{*}-L)$ into an integral is unchanged. What does
change are the expressions within the delta-functions and the determinant
of the fermion mass matrix. In particular, the positions of the vacua
are now determined by the conditions $D_{A}W+\widehat{K}_{A}W=0$,
where the second term is the correction due to warping. Thus, the
delta-functions must now read
\[
\delta^{2n+2}(Y_{A}+\widehat{K}_{A}X),
\]
and the quantities appearing in the fermion mass matrix now have to
incorporate warp corrections: $\left(D_{A}+\widehat{K}_{A}\right)\left(D_{B}+\widehat{K}_{B}\right)W$.
Note that at a vacuum we have the equivalence $\partial_{A}(D_{B}W+\widehat{K}_{B}W)\equiv\left(D_{A}+\widehat{K}_{A}\right)\left(D_{B}+\widehat{K}_{B}\right)W$,
and in general we will make use of similar equivalences in what follows.
We have:
\begin{eqnarray*}
D_{0}(D_{I}+\widehat{K}_{I})W & \equiv & Z_{I},\\
\left(D_{I}+\widehat{K}_{I}\right)\left(D_{J}+\widehat{K}_{J}\right)W & \equiv & \mathcal{F}_{IJK}\overline{Z}^{K}+\widehat{K}_{IJ}X+\widehat{K}_{J}Y_{I},\\
\left(D_{I}+\widehat{K}_{I}\right)\overline{\left(D_{J}+\widehat{K}_{J}\right)W} & \equiv & \left(\delta_{I\bar{J}}+\widehat{K}_{I\bar{J}}\right)\overline{X}.
\end{eqnarray*}
Upon integrating over the $Y_{A}$ we find the index density
\begin{eqnarray*}
\mu_{I} & = & 4^{n+1}|\det\eta|^{-1/2}\int_{-\infty}^{L_{*}}d\tilde{L}\int d^{2}X\, d^{2n}Z\,|\det g|\,\delta\left(\tilde{L}-\alpha|X|^{2}-|Z|^{2}\right)|X|^{2}\\
 & \times & \det\begin{pmatrix}\Xbar\mu_{I\bar{J}}-\frac{Z_{I}\Zbar_{\bar{J}}}{X} &  & \mathcal{F}_{IJK}\Zbar^{K}+\sigma_{IJ}X\\
\mathcal{\overline{F}}_{\overline{IJK}}Z^{\bar{K}}+\bar{\sigma}_{\overline{IJ}}\overline{X} &  & X\mu_{\bar{I}J}-\frac{\Zbar_{\bar{I}}Z_{J}}{\Xbar}
\end{pmatrix}
\end{eqnarray*}
where $\alpha=1-\widehat{K}_{I}\overline{\widehat{K}}^{I}$, $\mu_{I\bar{J}}=\delta_{I\bar{J}}+\widehat{K}_{I\bar{J}}$,
and $\sigma_{IJ}=\widehat{K}_{IJ}-\widehat{K}_{I}\widehat{K}_{J}$.

In order to compute this density, it proves helpful to consider the
special case of one complex modulus as well as the axio-dilaton. In
this particular case, we obtain the expression
\begin{equation}
\mu_{I}\propto\int_{-\infty}^{L_{*}}d\widetilde{L}\int d^{2}X\, d^{2n}Z\,|\det g|\,\delta\left(\widetilde{L}-\alpha|X|^{2}-|Z|^{2}\right)\left(|Z|^{4}+(\mu^{2}-|\sigma|^{2})|X|^{4}-(2\mu+|\mathcal{F}|^{2})|X|^{2}|Z|^{2}\right)
\end{equation}

Note, that we have eliminated a few terms that will integrate to zero
because they depend explicitly on the phases of $X,Z$. Far from the
conifold, $\alpha$ approaches $1$ since the warping corrections
can then be neglected. However, when one moves toward the conifold,
$\alpha$ gets progressively smaller until at some critical value
it equals zero, and then the warping correction drives $\alpha$ negative.
As long as $\alpha$ is positive, the tadpole delta function fixes
the range of integration so that $|X|$ and $|Z|$ lie on a finite
ellipse. Upon computing the integral, one therefore obtains a finite
value for the index. However, when $\alpha$ goes to zero, this ellipse
becomes increasingly stretched until for $\alpha=0$, the range of
integration for $|X|$ becomes unconstrained. At this point, the integral
above for the index density diverges. Then, as $\alpha$ goes negative,
this divergence persists as the ellipse turns into a hyperbola. Naively,
this suggests an infinite number of vacua within a finite disk surrounding
the conifold point. However, the more careful analysis taking account
of finite fluxes that we carry out below yields a finite result.

\subsection{Finite fluxes}

One major difference between the analysis above and numerical simulations
is the range of fluxes. In numerical simulations fluxes are necessarily
kept within a finite range, while in the derivation above, arbitrarily
large fluxes were included. To derive a theoretical distribution that
mirrors the effects seen in numerical studies, it is best to include
a bound on the fluxes in the analysis. This complicates the final
expression for the theoretical distribution but, of course, the finite
bound on fluxes is physically well motivated since the supergravity
approximation breaks down for large enough fluxes. In the absence
of warping, the finite range of fluxes does not lead to dramatic differences
from naively taking the bound to infinity, but as we will see, this
limit is more involved when warping is included.%
\footnote{By way of comparison, if we express the tadpole condition in the manner
of \cite{DouglasDenef}, it leads to an integral over a Gaussian-like
exponential factor $e^{-N\eta N/2}=e^{-|X|^{2}+|Y|^{2}-|Z|^{2}}$,
a damping term in the absence of warping due to the SUSY conditions
$\delta\left(Y_{A}\right)$. Warping modifies these conditions to
$Y_{A}+\widehat{K}_{A}X=0$, and so the argument of the exponential
is not negative definite, requiring an additional regulator bounding
the fluxes.%
}

Suppose that we bound our fluxes by the range $N_{i}\in[-\Lambda,\Lambda]$.
The $N_{a}$ and $X,Y,Z$ variables are related by
\begin{eqnarray}
X & = & N_{a}\Pi_{a}\\
Y_{A} & = & N_{a}D_{A}\Pi_{a}\\
Z_{I} & = & N_{a}D_{0}D_{I}\Pi_{a}
\end{eqnarray}
Here the $\Pi$'s are the periods of the rescaled holomorphic form,
as before. We would thus expect the ranges on $X,Y,Z$ to be moduli
dependent. Let's separate the phase and magnitude of $X,Y,Z$. Although
in principle, the ranges of the phases may have a complicated dependence
on both the moduli and the magnitudes $|X|,|Y|,|Z|$, we will neglect
this subtlety and suppose that they range over the usual $[0,2\pi]$.
As a result, we can easily integrate these variables out, leaving
us with the integrals over the magnitudes. We would expect to have
these range over the values
\begin{eqnarray}
|X| & \in & [0,\Lambda f_{X}(\xi,\tau)]\\
|Y_{A}| & \in & [0,\Lambda f_{Y}(\xi,\tau)]\\
|Z_{I}| & \in & [0,\Lambda f_{Z}(\xi,\tau)]
\end{eqnarray}
for particular $f_{X},f_{Y}$, and $f_{Z}$. Let's consider $f_{X}$.
The largest value that $|X|$ will take corresponds to the fluxes
$N_{a}$ taking one of their two extreme values of $\pm\Lambda$;
which of two possibilities maximizes $|X|$ depends on the near conifold
behavior of the periods. We must choose the eight signs for the eight
fluxes $N_{a}$ in such a way that we maximize the expression
\begin{equation}
f_{X}=\max\left(\left|\sum_{a}\pm\Pi_{a}(\xi,\tau)\right|\right)
\end{equation}

Since the periods are all finite in the near conifold limit, the $\xi$
dependence decouples. However, the value for $f_{X}$ will still be
$\tau$ dependent. As far as $f_{Y}$ and $f_{Z}$ are concerned,
the idea is the same except for the fact that the $\xi$ dependence
can't be neglected due to logarithmic divergences. In particular,
we find that
\begin{eqnarray}
f_{X} & = & f_{X}(\tau)\\
f_{Y} & = & f_{Y}^{1}(\tau)|1-f_{Y}^{2}(\tau)\log(\xi)|\\
f_{Z} & = & f_{Z}^{1}(\tau)|1-f_{Z}^{2}(\tau)\log(\xi)|
\end{eqnarray}

Now consider a fixed point in moduli space $\xi$ as well as a fixed
value for $\tau$. Then, the upper limits on these integrals will
involve particular constants multiplying the flux cutoff $\Lambda$.
Integrating over the variables $Y_{0}$ in the expression for the
index density, the delta function $\delta\left(Y_{0}\right)$ fixes
$Y_{0}=0$, leaving us with
\begin{eqnarray}
\mu_{I} & \propto & \int_{-\infty}^{L_{*}}d\widetilde{L}\int_{0}^{f_{X}\Lambda}|X|d|X|\int_{0}^{f_{Y}\Lambda}|Y|d|Y|\int_{0}^{f_{Z}\Lambda}|Z|d|Z|\,|\det g|\,\delta(\widetilde{L}-\alpha|X|^{2}-|Z|^{2})\nonumber \\
 & \times & \delta^{2}\left(Y_{1}+\widehat{K}_{\xi}X\right)\left(|Z|^{4}+\left(\mu^{2}-|\sigma|^{2}\right)|X|^{4}-\left(2\mu+|\mathcal{F}|^{2}\right)|X|^{2}|Z|^{2}\right)
\end{eqnarray}
The remaining delta function constraints come from the tadpole condition
and the supersymmetry condition $D_{\xi}W+\widehat{K}_{\xi}W=0$,
equivalent to $Y_{1}+\widehat{K}_{\xi}X=0$. Satisfying these constraints
will place complicated restrictions on the upper and lower bounds
of the remaining integrals. Let's first examine the region of integration
imposed by the supersymmetry constraint: 
\begin{itemize}
\item When $|X|$ is at its lower bound of 0, the constraint is trivial
to satisfy. Thus, the lower bound of $|X|$ is unchanged. 
\item However, when $|X|>0$, there will be points in the moduli space where
$\left|\widehat{K}_{\xi}X\right|>f_{Y}\Lambda$. At such points, the
delta function imposing the constraint $Y_{1}=-\widehat{K}_{\xi}X$
must vanish. We thus see that the upper bound of $|X|$ is restricted
in such cases to $f_{Y}\Lambda/\left|\widehat{K}_{\xi}\right|$. Solving
the delta function constraint for $Y_{1}$ requires that the upper
bound of the $|X|$ integral be taken to be $|X|_{\Lambda}=\textrm{min}\left(\Lambda f_{X},\Lambda f_{Y}/\left|\widehat{K}_{\xi}\right|\right)$. 
\end{itemize}
Note that in our scheme for bounding the fluxes, the upper limits
of integration for $|X|,|Y|$ and $|Z|$ all scale with the cutoff
$\Lambda$ in the same way. So, simply taking the limit as $\Lambda\rightarrow\infty$
won't affect the analysis. From our scheme's perspective, it is only
in the strictly infinite case where the naive divergence reappears
as discussed at the end of section \ref{sub:Incorporating-warping}.
(One could imagine more complicated schemes for bounding the fluxes,
treating $X,Y$, and $Z$ independently, allowing for a set of continuous
limits that recover the divergent results of the naive approach. Such
a scheme would increase the difficulty of relating the numerical and
theoretical analyses, as investigated in unwarped case in \cite{DouglasAshok,DouglasDenef,KachruTaxonomy}.

Given the new limits of integration on $|X|$, we can freely integrate
out the delta function fixing the value of $Y_{1}$:
\begin{eqnarray}
\mu_{I} & \propto & \int_{-\infty}^{L_{*}}d\widetilde{L}\int_{0}^{\left|X\right|_{\Lambda}}|X|d|X|\int_{0}^{f_{Z}\Lambda}|Z|d|Z|\,|\det g|\,\delta\left(\widetilde{L}-\alpha|X|^{2}-|Z|^{2}\right)\nonumber \\
 & \times & \left(|Z|^{4}+\left(\mu^{2}-|\sigma|^{2}\right)|X|^{4}-\left(2\mu+|\mathcal{F}|^{2}\right)|X|^{2}|Z|^{2}\right)
\end{eqnarray}
To simplify our notation, let's change variables to $u=|X|^{2}$ and
$v=|Z|^{2}$. The density can then be written as
\begin{equation}
\mu_{I}\propto\int_{-\infty}^{L_{*}}d\widetilde{L}\int_{0}^{u_{\Lambda}}du\int_{0}^{f_{Z}^{2}\Lambda^{2}}dv\,|\det g|\,\delta\left(\widetilde{L}-\alpha u-v\right)\left(v+\left(\mu^{2}-|\sigma|^{2}\right)u^{2}-\left(2\mu+|\mathcal{F}|^{2}\right)uv\right)\label{eq:IndexDensityFluxCutoff}
\end{equation}
where $u_{\Lambda}=|X|_{\Lambda}^{2}=\textrm{min}\left(\Lambda^{2}f_{X}^{2},\Lambda^{2}f_{Y}^{2}/\left|\widehat{K}_{\xi}\right|^{2}\right)$

It's useful to consider the two cases $\alpha>0$ and $\alpha<0$,
separately.

\begin{figure}[H]
\includegraphics[width=8cm]{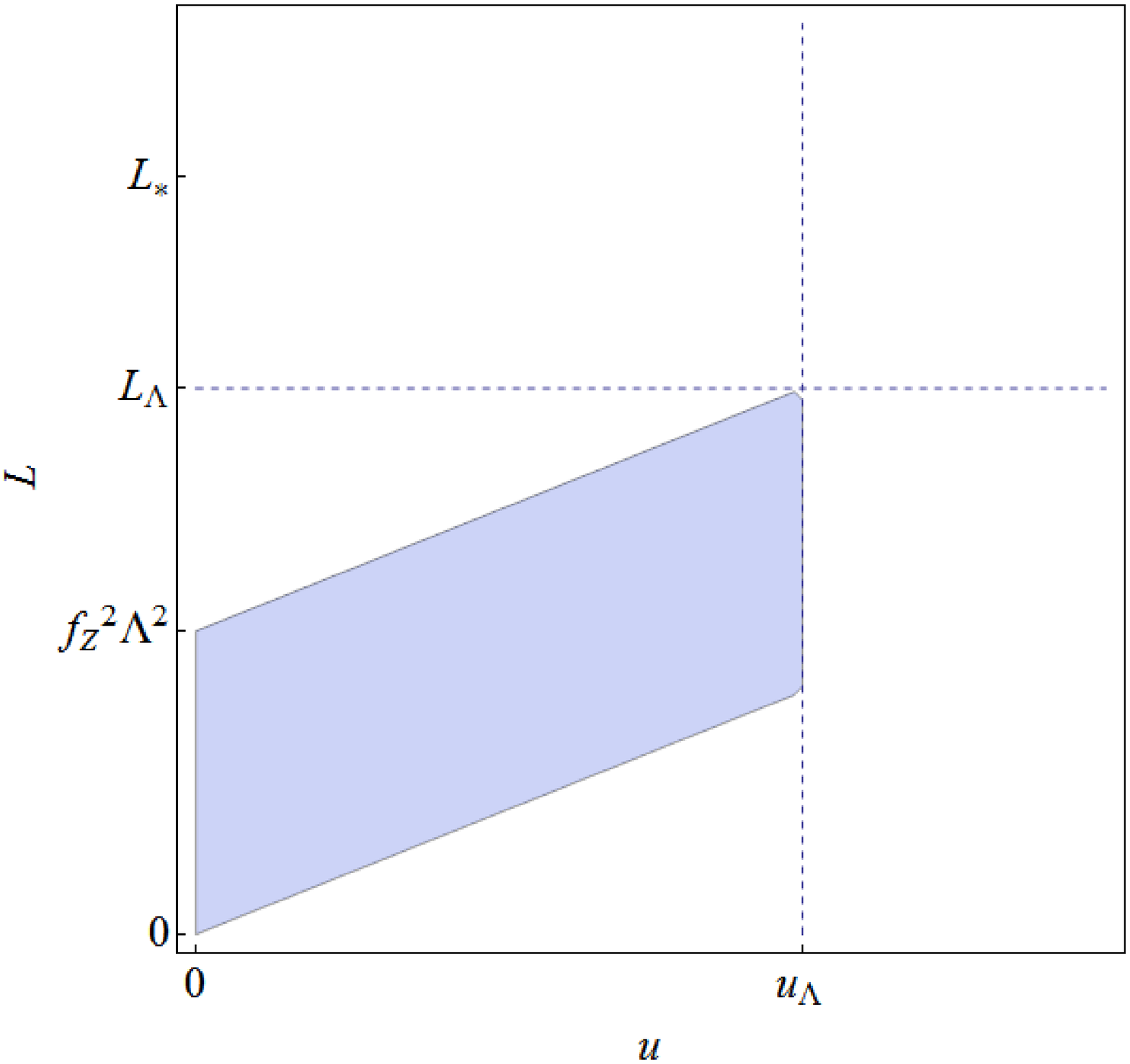}(a)\includegraphics[width=8cm]{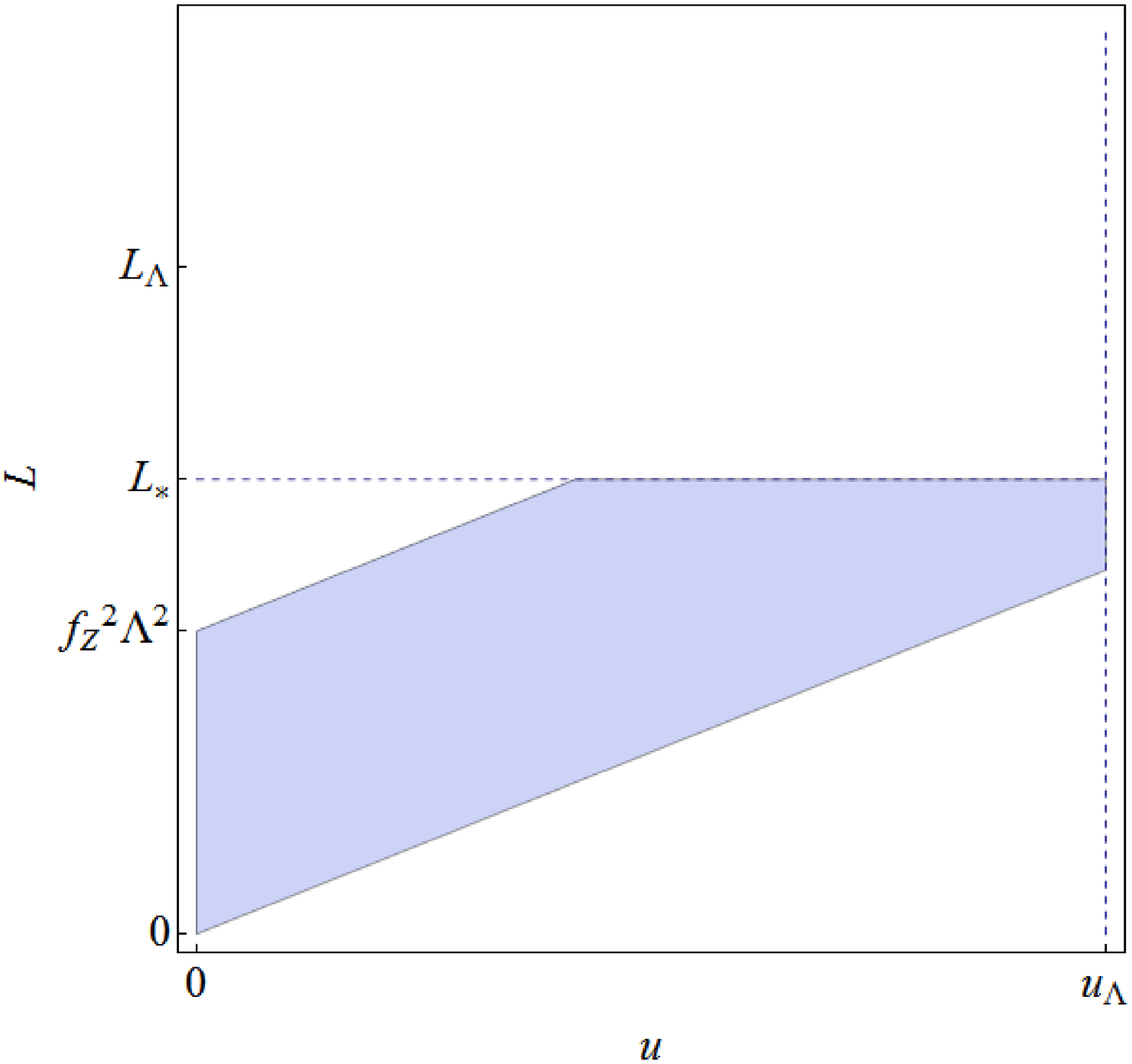}(b)

\includegraphics[width=8cm]{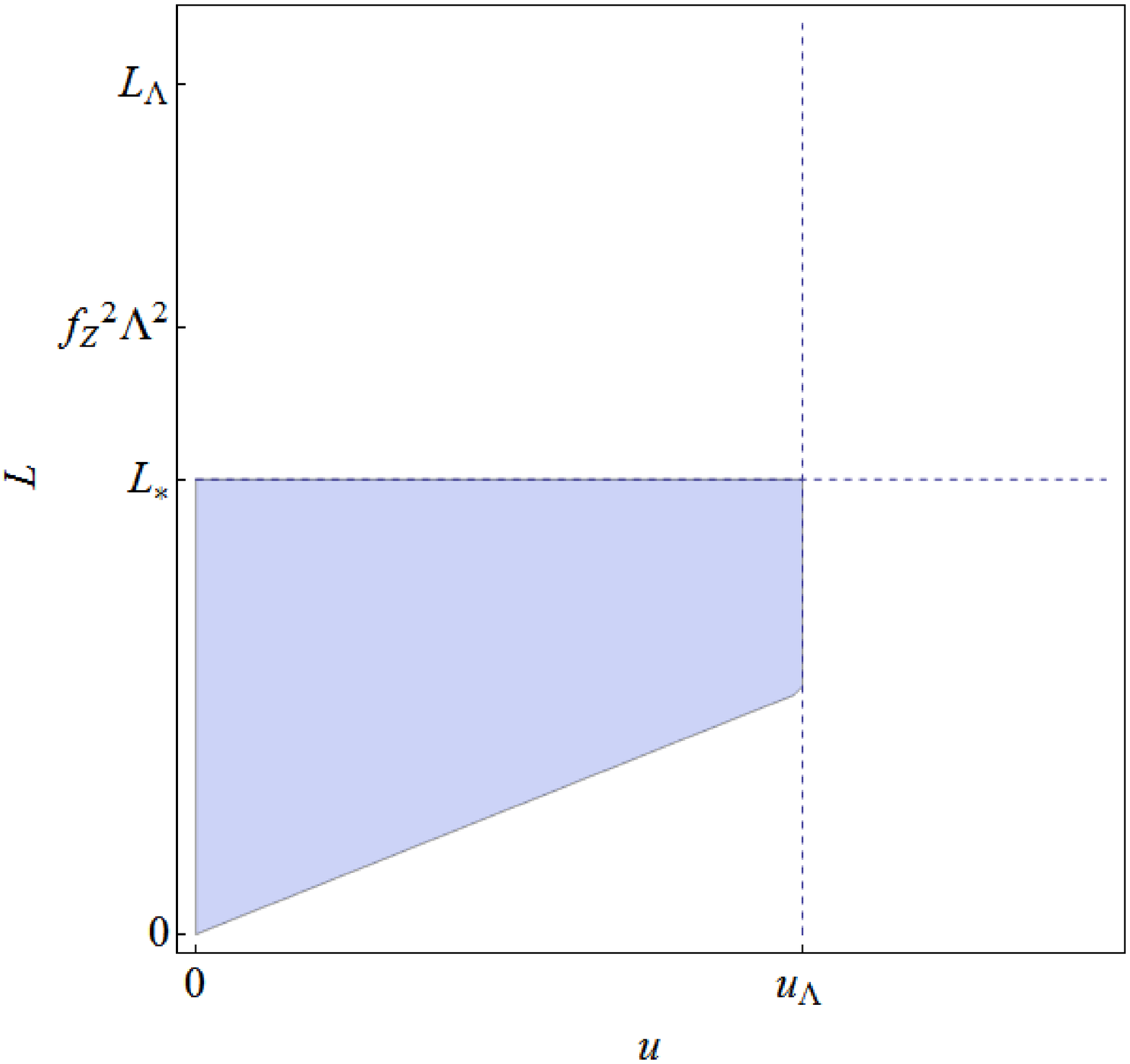}(c)\includegraphics[width=8cm]{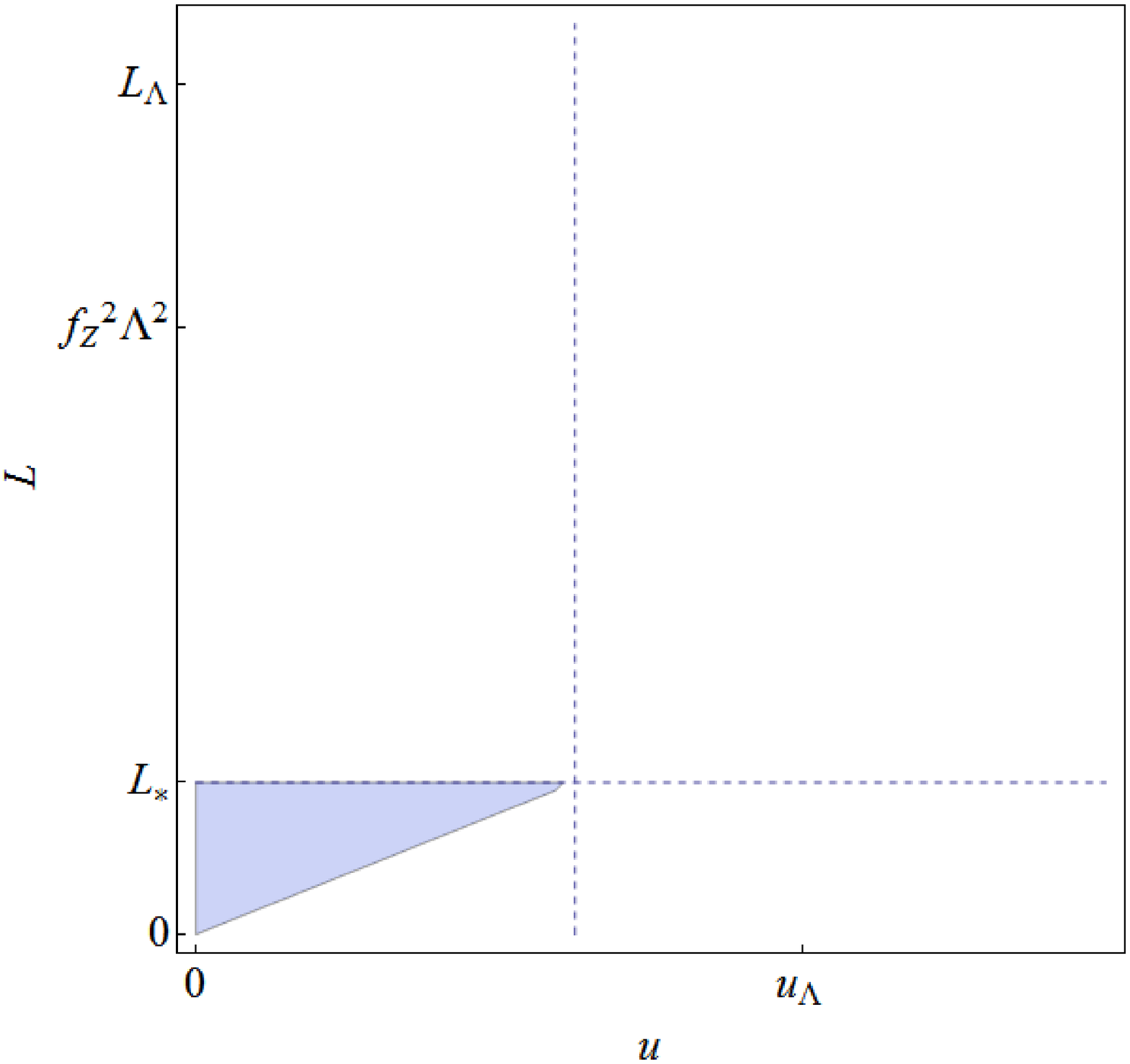}(d)

\caption{\label{fig:AlphaPlusRegions}Various possible regions of integration
for $\alpha>0$. In (a) the $L_{*}>L_{\Lambda},$ where $L_{\Lambda}=\alpha u_{\Lambda}+f_{Z}^{2}\Lambda^{2}$
and $u_{\Lambda}=\textrm{min}\left(\Lambda^{2}f_{X}^{2},\Lambda^{2}f_{Y}^{2}/\widehat{K}_{\xi}^{2}\right)$
so the region is cut off at $u_{\Lambda}$. In (b) $f_{Z}^{2}\Lambda^{2}<L_{*}<L_{\Lambda}$.
In (c), $L_{*}<f_{Z}^{2}\Lambda^{2}$, and $u_{\Lambda}<L_{*}/\alpha$.
Finally, (d) shows a region where $L_{*}<f_{Z}^{2}\Lambda^{2}$ and
$u_{\Lambda}>L_{*}/\alpha$.}
\end{figure}

\subsubsection{The case $\alpha>0$}

The delta function in (\ref{eq:IndexDensityFluxCutoff}) arising from
the tadpole condition is $\delta\left(\widetilde{L}-\alpha u-v\right)$.
This constrains the value of $v$ to be $\widetilde{L}-\alpha u$,
as well as constraining the region of integration on the $\widetilde{L}/u$-plane.
Let's first determine the lower bounds on $\widetilde{L}$ and $u$:
\begin{itemize}
\item The variables $u,v$ are positive or possibly zero and since $\alpha>0$,
then $\widetilde{L}\geq0$, fixing the lower bound of 0 for the $\widetilde{L}$
integral. 
\item Let $v_{\textrm{up}}=f_{Z}^{2}\Lambda^{2}$ be the upper bound on
$v$. If $v_{\textrm{up}}<\widetilde{L}$, then the delta function
forces the lower bound on $u$ to be $\left(\widetilde{L}-f_{Z}^{2}\Lambda^{2}\right)/\alpha$.
However, if $v_{\textrm{up}}>\widetilde{L}$ then the lower bound
on $u$ is 0. So in general we let the lower bound on $u$ be $u_{\textrm{down}}^{+}=\max\left(0,\frac{\widetilde{L}-f_{Z}^{2}\Lambda^{2}}{\alpha}\right)$. 
\end{itemize}
Now for the upper bounds: 
\begin{itemize}
\item If $L_{*}<\alpha u_{\Lambda}+f_{Z}^{2}\Lambda^{2}$, then it remains
the upper bound for the $\widetilde{L}$ integral. If on the contrary,
the inequality runs the other way, $L_{*}>\alpha u_{\Lambda}+f_{Z}^{2}\Lambda^{2}$,
then the constraint $\widetilde{L}-\alpha u-v$ cannot be satisfied
everywhere along the range $0<\widetilde{L}<L_{*}$, truncating this
range to $0<\widetilde{L}<\alpha u_{\Lambda}+f_{Z}^{2}\Lambda^{2}$
instead. So the upper bound on the $\widetilde{L}$ integral is
\[
L_{\textrm{up}}^{+}=\textrm{min}\left(L_{*},L_{\Lambda}\right)
\]
 where $L_{\Lambda}=\alpha u_{\Lambda}+f_{Z}^{2}\Lambda^{2}$. 
\item If, at a fixed $\widetilde{L}$, we had $\alpha u_{\Lambda}>\widetilde{L}$,
then since the lower bound on $v$ is 0, this places an upper bound
on $u$ of $\widetilde{L}/\alpha$. If the inequality is reversed,
then the upper bound on $u$ is $u_{\Lambda}$. So, in general the
upper bound on $u$ is $u_{\textrm{up}}^{+}=\min\left(u_{\Lambda},\widetilde{L}/\alpha\right)$. 
\end{itemize}
Various possible regions of integration in the $\widetilde{L}/u$-plane
are illustrated in figure \ref{fig:AlphaPlusRegions}.

Using the bounds described above and integrating over $v$ yields
\begin{align*}
\mu_{I}^{+} & \propto\int_{0}^{L_{\textrm{up}}^{+}}d\widetilde{L}\int_{u_{\textrm{down}}^{+}\left(\widetilde{L}\right)}^{u_{\textrm{up}}^{+}\left(\widetilde{L}\right)}du|\det g|\left(\left(\widetilde{L}-\alpha u\right)^{2}+\beta u^{2}+\gamma u\left(\widetilde{L}-\alpha u\right)\right)
\end{align*}
where and $\beta=\mu^{2}-|\sigma|^{2}$ and $\gamma=-2\mu-|\mathcal{F}|^{2}$.
Then, expanding everything out and integrating over $u$, we obtain
\begin{align*}
\mu_{I}^{+} & \propto\int_{0}^{L_{\textrm{up}}^{+}}d\widetilde{L}|\det g|\Bigg(\widetilde{L}^{2}\left(u_{\textrm{up}}^{+}-u_{\textrm{down}}^{+}\right)+\frac{\gamma-2\alpha}{2}\widetilde{L}\left(\left(u_{\textrm{up}}^{+}\right)^{2}-\left(u_{\textrm{down}}^{+}\right)^{2}\right)\\
 & +\frac{\alpha^{2}+\beta-\alpha\gamma}{3}\left(\left(u_{\textrm{up}}^{+}\right)^{3}-\left(u_{\textrm{down}}^{+}\right)^{3}\right)\Bigg)
\end{align*}
where the $\widetilde{L}$ dependence of $u_{\textrm{up}}$ and $u_{\textrm{down}}$
has been suppressed in the last line.

In order to integrate over $\widetilde{L}$, we must separate the
integral above into two parts since $u_{\textrm{up}}^{+}$ and $u_{\textrm{down}}^{+}$
are different functions of $\widetilde{L}$. Let $\mathscr{I}_{\textrm{up}}^{+}$
be the portion of the integral involving terms containing powers of
$u_{\textrm{up}}^{+}$ and $\mathscr{I}_{\textrm{down}}^{+}$ be the
portion of the integral containing $u_{\textrm{down}}^{+}$. Note
that we remove the $\left|\det g\right|$ factor from these integrals.
Focusing first on $\mathscr{I}_{\textrm{up}}^{+}$ we see that for%
\footnote{Note that we could have considered $u_{\textrm{up}}^{+}\left(L_{\textrm{up}}^{+}\right)$
instead of $u_{\textrm{up}}^{+}\left(L_{*}\right)$ as the upper part
of the interval. However, recall that $L_{\textrm{up}}^{+}$ is the
smaller of either $L_{*}$ or $L_{\Lambda}$. If $L_{*}>L_{\Lambda}$,
$u_{\textrm{up}}^{+}\left(L_{\textrm{up}}^{+}\right)=u_{\textrm{up}}^{+}\left(L_{\Lambda}\right)=u_{\Lambda}$
since from the definition of $L_{\Lambda}$, the intequality $u_{\Lambda}<L_{\Lambda}/\alpha$
always holds.%
} $0<\widetilde{L}/\alpha<u_{\textrm{up}}^{+}\left(L_{*}\right)$,
we can replace instances of $u_{\textrm{up}}^{+}\left(\widetilde{L}\right)$
in the integral with $\widetilde{L}/\alpha$, while if $u_{\textrm{up}}^{+}\left(L_{*}\right)<\widetilde{L}/\alpha$,
then $u_{\textrm{up}}^{+}=u_{\Lambda}$, which is independent of $\widetilde{L}$.
So $\mathscr{I}_{\textrm{up}}^{+}$ splits into integrals over the
two regions:
\begin{eqnarray}
\mathscr{I}_{\textrm{up}}^{+} & = & \int_{0}^{\alpha u_{\textrm{up}}^{+}\left(L_{*}\right)}d\widetilde{L}\left(\frac{1}{\alpha}+\frac{\gamma-2\alpha}{2\alpha^{2}}+\frac{\alpha^{2}+\beta-\alpha\gamma}{3\alpha^{3}}\right)\widetilde{L}^{3}\nonumber \\
 & + & \int_{\alpha u_{\textrm{up}}^{+}\left(L_{*}\right)}^{L_{\textrm{up}}^{+}}d\tilde{L}\left(\widetilde{L}^{2}u_{\Lambda}+\frac{\gamma-2\alpha}{2}\widetilde{L}u_{\Lambda}^{2}+\frac{\alpha^{2}+\beta-\alpha\gamma}{3}u_{\Lambda}^{3}\right)\label{eq:IupBeforeIntegral}
\end{eqnarray}
Integrating yields
\begin{align}
\mathscr{I}_{\textrm{up}}^{+} & =\left(\frac{1}{\alpha}+\frac{\gamma-2\alpha}{2\alpha^{2}}+\frac{\alpha^{2}+\beta-\alpha\gamma}{3\alpha^{3}}\right)\frac{\alpha^{4}\left(u_{\textrm{up}}^{+}\left(L_{*}\right)\right)^{4}}{4}\nonumber \\
 & +\frac{\left(\left(L_{\textrm{up}}^{+}\right)^{3}-\alpha^{3}\left(u_{\textrm{up}}^{+}\left(L_{*}\right)\right)^{3}\right)u_{\Lambda}}{3}+\frac{\gamma-2\alpha}{4}\left(\left(L_{\textrm{up}}^{+}\right)^{2}-\alpha^{2}\left(u_{\textrm{up}}^{+}\left(L_{*}\right)\right)^{3}\right)u_{\Lambda}^{2}\label{eq:IupAfterIntegral}\\
 & +\frac{\alpha^{2}+\beta-\alpha\gamma}{3}\left(L_{\textrm{up}}^{+}-\alpha u_{\textrm{up}}^{+}\left(L_{*}\right)\right)u_{\Lambda}^{3}
\end{align}

For the integral $\mathscr{I}_{\textrm{down}}^{+}$, we consider the
regions $0<\widetilde{L}<f_{Z}^{2}\Lambda$ and $f_{Z}^{2}\Lambda<\widetilde{L}$.
In the first case, $u_{\textrm{down}}^{+}=0$ in which case this entire
portion of the integral vanishes, while in the second, $u_{\textrm{down}}^{+}=\left(\widetilde{L}-f_{Z}^{2}\Lambda\right)/\alpha$.
If $L_{*}<f_{Z}^{2}\Lambda^{2}$ then the entirety of $\mathscr{I}_{\textrm{down}}^{+}=0$,
so we have
\begin{align*}
\mathscr{I}_{\textrm{down}}^{+} & =-\theta\left(L_{*}-f_{Z}^{2}\Lambda^{2}\right)\int_{f_{Z}^{2}\Lambda^{2}}^{L_{\textrm{up}}^{+}}d\widetilde{L}\Bigg(\frac{1}{\alpha}\left(\widetilde{L}^{3}-f_{Z}^{2}\Lambda^{2}\widetilde{L}^{2}\right)+\frac{\gamma-2\alpha}{2\alpha^{2}}\left(\widetilde{L}^{3}-2f_{Z}^{2}\Lambda^{2}\widetilde{L}^{2}+f_{Z}^{4}\Lambda^{4}\widetilde{L}\right)\\
 & +\frac{\alpha^{2}+\beta-\alpha\gamma}{3\alpha^{3}}\left(\widetilde{L}^{3}-3f_{Z}^{2}\Lambda^{2}\widetilde{L}^{2}+3f_{Z}^{4}\Lambda^{4}\widetilde{L}-f_{Z}^{6}\Lambda^{6}\right)\Bigg)\\
 & =-\theta\left(L_{*}-f_{Z}^{2}\Lambda^{2}\right)\int_{f_{Z}^{2}\Lambda^{2}}^{L_{\textrm{up}}^{+}}d\widetilde{L}\Bigg(\left(\frac{1}{\alpha}+\frac{\gamma-2\alpha}{2\alpha^{2}}+\frac{\alpha^{2}+\beta-\alpha\gamma}{3\alpha^{3}}\right)\widetilde{L}^{3}\\
 & -f_{Z}^{2}\Lambda^{2}\left(\frac{1}{\alpha}+\frac{\gamma-2\alpha}{\alpha^{2}}+\frac{\alpha^{2}+\beta-\alpha\gamma}{\alpha^{3}}\right)\widetilde{L}^{2}\\
 & +f_{Z}^{4}\Lambda^{4}\left(\frac{\gamma-2\alpha}{2\alpha^{2}}+\frac{\alpha^{2}+\beta-\alpha\gamma}{\alpha^{3}}\right)\widetilde{L}-f_{Z}^{6}\Lambda^{6}\frac{\alpha^{2}+\beta-\alpha\gamma}{3\alpha^{3}}\Bigg)
\end{align*}
Integrating yields
\begin{align*}
\mathscr{I}_{\textrm{down}}^{+} & =-\theta\left(L_{*}-f_{Z}^{2}\Lambda^{2}\right)\Bigg(f_{Z}^{8}\Lambda^{8}\left(\frac{1}{12\alpha}-\frac{\gamma-2\alpha}{24\alpha^{2}}+\frac{\alpha^{2}+\beta-\alpha\gamma}{12\alpha^{3}}\right)\\
 & -f_{Z}^{6}\Lambda^{6}\frac{\alpha^{2}+\beta-\alpha\gamma}{3\alpha^{3}}L_{\textrm{up}}^{+}+f_{Z}^{4}\Lambda^{4}\left(\frac{\gamma-2\alpha}{4\alpha^{2}}+\frac{\alpha^{2}+\beta-\alpha\gamma}{2\alpha^{3}}\right)\left(L_{\textrm{up}}^{+}\right)^{2}\\
 & -\frac{1}{3}f_{Z}^{2}\Lambda^{2}\left(\frac{1}{\alpha}+\frac{\gamma-2\alpha}{\alpha^{2}}+\frac{\alpha^{2}+\beta-\alpha\gamma}{\alpha^{3}}\right)\left(L_{\textrm{up}}^{+}\right)^{3}+\left(\frac{1}{4\alpha}+\frac{\gamma-2\alpha}{8\alpha^{2}}+\frac{\alpha^{2}+\beta-\alpha\gamma}{12\alpha^{3}}\right)\left(L_{\textrm{up}}^{+}\right)^{4}\Bigg)
\end{align*}

Notice that when one ignores warping and the finite fluxes, $\alpha=1,\widehat{K}_{\xi}=0$,
and $\Lambda\rightarrow\infty$, implying $\beta=1$, $\gamma=-2-\left|\mathcal{F}\right|^{2}$,
$u_{\textrm{up}}^{+}\left(L_{*}\right)=L_{*}$, and $L_{\textrm{up}}^{+}=L_{*}$.
In this case, we must go back to the expression (\ref{eq:IupBeforeIntegral})
and note that the second integral in that expression vanishes since
the lower and upper bound of integration are both $L_{*}$. Furthermore
the integral $\mathscr{I}_{\textrm{down}}^{+}$ vanishes due to the
$\theta$-function prefactor. The index density in the unwarped case
is thus
\begin{equation}
\mu_{I}^{\text{Unwarped}}(\xi,\tau)=\left|\det g\right|\mathscr{I}_{\textrm{up}}^{+}=\left|\det g\right|\frac{L_{*}^{4}}{4}\left(\frac{6+3\gamma-6+2+2\beta-2\gamma}{6}\right)=\left|\det g\right|\frac{L_{*}^{4}}{4!}(2-|\mathcal{F}|^{2})
\end{equation}
This precise combination gives us the curvature tensor as argued in
\cite{DouglasAshok,DouglasDenef}. So, our expression reduces to the
correct form in the unwarped, infinite flux case. We now turn our
attention to the case where $\alpha<0$.

\subsubsection{The case $\alpha<0$}

Once again, we first establish the lower and upper bounds on $\widetilde{L}$
and $u$: 
\begin{itemize}
\item Suppose we are at the lower bound on $v$, namely $v=0$. In this
case, the tadpole constraint $\widetilde{L}-\alpha u-v=0$ tells us
that the lower bound attained by $\widetilde{L}$ is $L_{\textrm{down}}^{-}=\alpha u_{\Lambda}$.
Note that this is negative. 
\item Consider some fixed $\widetilde{L}\leq f_{Z}^{2}\Lambda^{2}$; If
$\widetilde{L}>0$, then there is always a $v=\widetilde{L}$ to cancel
it, and the lower bound for $u$ in this case is 0. However, if $\widetilde{L}<0$,
the fact that $v\geq0$ implies that for the constraint to hold, we
need the lower bound for $u$ to be $\widetilde{L}/\alpha$. So in
general, the lower bound for $u$ is $u_{\textrm{down}}^{-}=\max\left(0,\widetilde{L}/\alpha\right)$. 
\item By similar reasoning to the previous case, if $L_{*}<f_{Z}^{2}\Lambda^{2}$,
then it may remain the upper bound on $\widetilde{L}$. However, if
$L_{*}>f_{Z}^{2}\Lambda^{2}$, then the upper bound on $\widetilde{L}$
becomes $f_{Z}^{2}\Lambda^{2}$. So in general, the upper bound on
$\widetilde{L}$ is $L_{\textrm{up}}^{-}=\min(L_{*},f_{Z}^{2}\Lambda^{2})$. 
\item Consider again a fixed $\widetilde{L}$, and suppose $v$ is at its
upper bound of $f_{Z}^{2}\Lambda^{2}$. If $\alpha u_{\Lambda}>\widetilde{L}-f_{Z}^{2}\Lambda^{2}$,
then the upper bound of $u$ must be truncated to $\left(\widetilde{L}-f_{Z}^{2}\Lambda^{2}\right)/\alpha$.
Otherwise, if $\alpha u_{\Lambda}<\widetilde{L}-f_{Z}^{2}\Lambda^{2}$,
then the upper bound on $u$ remains $u_{\Lambda}$. In general then,
$u_{\textrm{up}}^{-}=\min\left(u_{\Lambda},\left(\widetilde{L}-f_{Z}^{2}\Lambda^{2}\right)/\alpha\right)$. 
\end{itemize}
Given these bounds on $u$ and $\widetilde{L}$, we may now integrate
over $v$, eliminating the tadpole delta function to get
\begin{equation}
\mu_{I}^{-}\propto\int_{L_{\textrm{down}}^{-}}^{L_{\textrm{up}}^{-}}d\widetilde{L}\int_{u_{\textrm{down}}^{-}\left(\widetilde{L}\right)}^{u_{\textrm{up}}^{-}\left(\widetilde{L}\right)}du\,|\det g|\,\left(\widetilde{L}^{2}+(\gamma-2\alpha)\widetilde{L}u+\left(\alpha^{2}+\beta-\alpha\gamma\right)u^{2}\right)
\end{equation}
Carrying out the $u$ integration yields
\begin{eqnarray}
\mu_{I}^{-} & \propto & \int_{L_{\textrm{down}}^{-}}^{L_{\textrm{up}}^{-}}d\widetilde{L}|\det g|\,\Bigg(\widetilde{L}^{2}\left(u_{\textrm{up}}^{-}-u_{\textrm{down}}^{-}\right)+\widetilde{L}\left(\frac{\gamma-2\alpha}{2}\right)\left(\left(u_{\textrm{up}}^{-}\right)^{2}-\left(u_{\textrm{down}}^{-}\right)^{2}\right)\nonumber \\
 & + & \frac{\alpha^{2}+\beta-\alpha\gamma}{3}\left(\left(u_{\textrm{up}}^{-}\right)^{3}-\left(u_{\textrm{down}}^{-}\right)^{3}\right)\Bigg)
\end{eqnarray}
where we have suppressed the $\widetilde{L}$ dependence of $u_{\textrm{up}}^{-}$,
and $u_{\textrm{down}}^{-}$.

As before, split the integral into two parts, $\mathscr{I}_{\textrm{up}}^{-}$
and $\mathscr{I}_{\textrm{down}}^{-}$, involving just the $u_{\textrm{up}}^{-}$
and $u_{\textrm{down}}^{-}$ parts, respectively. To compute $\mathscr{I}_{\textrm{up}}^{-}$
we consider two cases: 
\begin{itemize}
\item Suppose $L_{*}<L_{\Lambda}$, where we recall $L_{\Lambda}=\alpha u_{\Lambda}+f_{Z}^{2}\Lambda^{2}$.
Note that since $\alpha<0$, we have that $L_{\Lambda}<f_{Z}^{2}\Lambda^{2}$,
and thus, $L_{\textrm{up}}^{-}=L_{*}$ in this case. We also see that
$u_{\textrm{up}}^{-}=u_{\Lambda}$, and so in this case, the integral
$\mathscr{I}_{\textrm{up}}^{-}$ is simply
\[
\mathscr{I}_{\textrm{up}}^{-}=\int_{L_{\textrm{down}}^{-}}^{L_{*}}d\widetilde{L}\Bigg(\widetilde{L}^{2}u_{\Lambda}+\widetilde{L}\left(\frac{\gamma-2\alpha}{2}\right)u_{\Lambda}^{2}+\frac{\alpha^{2}+\beta-\alpha\gamma}{3}u_{\Lambda}^{3}\Bigg)
\]

\item Suppose that $L_{*}>L_{\Lambda}$. In this case, for $\widetilde{L}<L_{\Lambda}$,
$u_{\textrm{up}}^{-}=u_{\Lambda}$ as before, but when $\widetilde{L}>L_{\Lambda}$
we have $u_{\textrm{up}}^{-}=\left(\widetilde{L}-f_{Z}^{2}\Lambda^{2}\right)/\alpha$.
So the integral splits into two parts
\begin{eqnarray}
\mathscr{I}_{\textrm{up}}^{-} & = & \int_{L_{\textrm{down}}^{-}}^{L_{\Lambda}}d\widetilde{L}\Bigg(\widetilde{L}^{2}u_{\Lambda}+\widetilde{L}\left(\frac{\gamma-2\alpha}{2}\right)u_{\Lambda}^{2}+\frac{\alpha^{2}+\beta-\alpha\gamma}{3}u_{\Lambda}^{3}\Bigg)\nonumber \\
 & + & \int_{L_{\Lambda}}^{L_{\textrm{up}}^{-}}d\widetilde{L}\Bigg(\widetilde{L}^{2}\left(\frac{\widetilde{L}-f_{Z}^{2}\Lambda^{2}}{\alpha}\right)+\widetilde{L}\left(\frac{\gamma-2\alpha}{2}\right)\left(\frac{\widetilde{L}-f_{Z}^{2}\Lambda^{2}}{\alpha}\right)^{2}\nonumber \\
 & + & \frac{\alpha^{2}+\beta-\alpha\gamma}{3}\left(\frac{\widetilde{L}-f_{Z}^{2}\Lambda^{2}}{\alpha}\right)^{3}\Bigg)
\end{eqnarray}

\end{itemize}
These two expressions can be joined if we introduce $L_{\textrm{mid}}=\min\left(L_{*},L_{\Lambda}\right)$:
\begin{eqnarray}
\mathscr{I}_{\textrm{up}}^{-} & = & \int_{L_{\textrm{down}}^{-}}^{L_{\textrm{mid}}}d\widetilde{L}\Bigg(\widetilde{L}^{2}u_{\Lambda}+\widetilde{L}\left(\frac{\gamma-2\alpha}{2}\right)u_{\Lambda}^{2}+\frac{\alpha^{2}+\beta-\alpha\gamma}{3}u_{\Lambda}^{3}\Bigg)\nonumber \\
 & + & \int_{L_{\textrm{mid}}}^{L_{\textrm{up}}^{-}}d\widetilde{L}\Bigg(\widetilde{L}^{2}\left(\frac{\widetilde{L}-f_{Z}^{2}\Lambda^{2}}{\alpha}\right)+\left(\frac{\gamma-2\alpha}{2}\right)\widetilde{L}\left(\frac{\widetilde{L}-f_{Z}^{2}\Lambda^{2}}{\alpha}\right)^{2}\nonumber \\
 & + & \frac{\alpha^{2}+\beta-\alpha\gamma}{3}\left(\frac{\widetilde{L}-f_{Z}^{2}\Lambda^{2}}{\alpha}\right)^{3}\Bigg)
\end{eqnarray}
 The integral in the second line above vanishes if $L_{\textrm{mid}}=L_{*}$,
since in that case $L_{\textrm{up}}^{-}$ also is $L_{*}$. Carrying
out the integral yields (after plugging in $L_{\textrm{down}}^{-}=\alpha u_{\Lambda}$)
\begin{eqnarray*}
\mathscr{I}_{\textrm{up}}^{-} & = & \frac{u_{\Lambda}}{3}\bigg(L_{\textrm{mid}}^{3}-\alpha^{3}u_{\Lambda}^{3}\bigg)+\frac{\gamma-2\alpha}{4}u_{\Lambda}^{2}\bigg(L_{\textrm{mid}}^{2}-\alpha^{2}u_{\Lambda}^{2}\bigg)+\frac{\alpha^{2}+\beta-\alpha\gamma}{3}u_{\Lambda}^{3}\left(L_{\textrm{mid}}-\alpha u_{\Lambda}\right)\\
 & + & \frac{1}{4}\left(\frac{1}{\alpha}+\frac{\gamma-2\alpha}{2\alpha^{2}}+\frac{\alpha^{2}+\beta-\alpha\gamma}{3\alpha^{3}}\right)\left(\left(L_{\textrm{up}}^{-}\right)^{4}-L_{\textrm{mid}}^{4}\right)\\
 & - & \frac{f_{Z}^{2}\Lambda^{2}}{3}\left(\frac{1}{\alpha}+\frac{\gamma-2\alpha}{\alpha^{2}}+\frac{\alpha^{2}+\beta-\alpha\gamma}{\alpha^{3}}\right)\left(\left(L_{\textrm{up}}^{-}\right)^{3}-L_{\textrm{mid}}^{3}\right)\\
 & + & \frac{f_{Z}^{4}\Lambda^{4}}{2}\left(\frac{\gamma-2\alpha}{2\alpha^{2}}+\frac{\alpha^{2}+\beta-\alpha\gamma}{\alpha^{3}}\right)\left(\left(L_{\textrm{up}}^{-}\right)^{2}-L_{\textrm{mid}}^{2}\right)\\
 & - & f_{Z}^{6}\Lambda^{6}\left(\frac{\alpha^{2}+\beta-\alpha\gamma}{3\alpha^{3}}\right)\left(L_{\textrm{up}}^{-}-L_{\textrm{mid}}\right)
\end{eqnarray*}
The integral $\mathscr{I}_{\textrm{down}}^{-}$ vanishes when $\widetilde{L}>0$
since in that case $u_{\textrm{down}}^{-}=0$. Thus, the only region
that contributes is where $L_{\textrm{down}}^{-}\leq\widetilde{L}\leq0$,
in which $u_{\textrm{down}}^{-}=\widetilde{L}/\alpha$. We have,
\[
\mathscr{I}_{\textrm{down}}^{-}=-\int_{L_{\textrm{down}}^{-}}^{0}d\widetilde{L}\left(\frac{1}{\alpha}\tilde{L}^{3}+\frac{\gamma-2\alpha}{2\alpha^{2}}\tilde{L}^{3}+\frac{\alpha^{2}+\beta-\alpha\gamma}{3\alpha^{3}}\tilde{L}^{3}\right)
\]
 which gives
\[
\mathscr{I}_{\textrm{down}}^{-}=\frac{1}{4}\left(\frac{1}{\alpha}+\frac{\gamma-2\alpha}{2\alpha^{2}}+\frac{\alpha^{2}+\beta-\alpha\gamma}{3\alpha^{3}}\right)\alpha^{4}u_{\Lambda}^{4}
\]
 where we have again used $L_{\textrm{down}}^{-}=\alpha u_{\Lambda}$.

The full index density is thus
\begin{equation}
\mu_{I}(\xi,\tau)/\det g=\left(\mathscr{I}_{\textrm{up}}^{+}+\mathscr{I}_{\textrm{down}}^{+}\right)\theta(\alpha)+\left(\mathscr{I}_{\textrm{up}}^{-}+\mathscr{I}_{\textrm{down}}^{-}\right)\theta(-\alpha)
\end{equation}

In the unwarped, infinite flux case where a consise geometric result
is obtained, one can integrate out the axio-dilaton to obtain an effective
density only in terms of the complex moduli. However, in our case
this type of integration proves intractable. As a result we will when
comparing with simulations have to fix a value of the axio-dilaton
and compare the un-integrated form of our density.

\section{Numerical Vacuum Statistics\label{sec:Numerical-Vacuum-Statistics}}

To perform a numerical study of the distribution of vacua in moduli
space near the conifold point, we will randomly choose appropriate
fluxes $F=(F_{0},F_{1},F_{2},F_{3})$ and $H=(H_{0},H_{1},H_{2},H_{3})$
and then solve the conditions $D_{\tau}W=0$ and $D_{\xi}W=0$ for
the moduli space coordinate $\xi$. Here $W=N_{i}\Pi_{i}$ is the
superpotential, and $N$ is an 8-vector whose first four components
are those of $F$ and last four are those of $H$. We work in a basis
such that the vector of 4-fold periods $\Pi=\left(\Sigma,\tau\Sigma\right)$,
where $\Sigma$ is the vector of periods on the 3-fold. Near the conifold
point, the vector of 3-fold periods takes the form:
\begin{equation}
\Sigma=\sum_{n=0}^{\infty}a_{n}\xi^{n}+b\xi\log(-i\xi),
\end{equation}
 where the $a_{n}$ and $b$ are constant vectors associated with
the expansion of the periods. Note that in the case of a single complex
modulus, the vector $b=(0,0,0,b^{0})$, since only $\Sigma_{0}$ has
non-trivial logarithmic behavior near the conifold. Also, the local
coordinate around the conifold point is proportional to $\Sigma_{3}$,
which implies that the vector $a_{0}=(0,a_{0}^{2},a_{0}^{1},a_{0}^{0})$.

\subsection{Unwarped Analysis}

The unwarped Kähler potential is
\begin{equation}
e^{-K}=-i\overline{\Sigma}\cdot Q\cdot\Sigma=-i\left((\abar_{n}\cdot Q\cdot a_{m})\xibar^{n}\xi^{m}+(\bbar\cdot Q\cdot a_{m})\xi^{m}\xibar\log(i\overline{\xi})+(\abar_{n}\cdot Q\cdot b)\xibar^{n}\xi\log(-i\xi)\right),
\end{equation}
 where the term proportional to $\bbar\cdot Q\cdot b$ has been dropped
since given $b$ and $\eta$ it vanishes.

For SUSY vacua in the unwarped case
\begin{equation}
D_{\xi}W=N\cdot(\partial_{\xi}\Pi+\Pi K_{\xi})=0.
\end{equation}
 Keeping logarithmic and constant terms gives
\begin{equation}
(F-\tau H)\cdot\left(a_{1}+b\left(\log(-i\xi)+1\right)\right)-(F-\tau H)\cdot a_{0}\frac{\abar_{0}\cdot Q\cdot a_{1}}{\abar_{0}\cdot Q\cdot a_{0}}=0.
\end{equation}
 where the fact that $\abar_{0}\cdot Q\cdot b=0$ has been used to
simplify the expression. This is an equation of the form
\begin{equation}
\mathcal{A}+\mathcal{B}\log(-i\xi)=0,
\end{equation}
 with
\begin{eqnarray}
\mathcal{A} & = & \frac{1}{c}(F-\tau H)\cdot b(\abar_{0}\cdot Q\cdot a_{0})+(F-\tau H)\cdot a_{1}(\abar_{0}\cdot Q\cdot a_{0})-(F-\tau H)\cdot a_{0}(\abar_{1}\cdot Q\cdot a_{0})\\
\mathcal{B} & = & (F-\tau H)\cdot b(\abar_{0}\cdot Q\cdot a_{0}).
\end{eqnarray}

The leading-order constraints arising from requiring $D_{\tau}W=0$
are
\begin{equation}
\tau=\frac{F\cdot\overline{\Sigma}}{H\cdot\overline{\Sigma}}=\frac{F\cdot\abar_{0}}{H\cdot\abar_{0}}.\label{eq:tau}
\end{equation}
 This implies that
\begin{equation}
F-\tau H=\frac{(H\cdot\abar_{0})F-(F\cdot\abar_{0})H}{H\cdot\abar_{0}}
\end{equation}

Before considering the effects of warp corrections, it's worth determining
how close to the conifold vacua may be found in the unwarped scenario.
The $D_{\xi}W=0$ constraint implies that $|c\xi|$ is exponentially
suppressed by the ratio of $|\mathcal{A}/\mathcal{B}|$, so if $|\mathcal{A}|$
is even just a couple of orders of magnitude greater than $|\mathcal{B}|$,
we should expect to see vacua on the order of $10^{-100}$ units away
from the conifold point---indeed, this has been observed in previous
studies. In order for $|\mathcal{A}|$ to differ appreciably from
$|\mathcal{B}|$ the quantity $|(F-\tau H)\cdot b|$ should be relatively
small compared to $|(F-\tau H)\cdot a_{0}|$ or $|(F-\tau H)\cdot a_{1}|$.
Using the form of the vector $b$ above, this indicates that the fluxes
through the collapsing cycle, $F_{3}$ and $H_{3}$ should be small
relative to some of the other fluxes.

\subsection{Warped Analysis}

Introducing warping leads to the corrections (\ref{WarpedKahler})
and (\ref{WarpedKahlerDerivative}) to the Kähler potential and its
derivative. The modification to the near-conifold SUSY vacuum condition
is then
\begin{equation}
D_{\xi}W\longrightarrow D_{\xi}W+3C_{w}\frac{\xibar^{1/3}}{\xi^{2/3}}N\cdot\Pi.
\end{equation}
 Now, assuming that $C_{w}$ is small (i.e. the volume of the 3-fold
is large) these new terms will matter only close to $\xi=0$. The
SUSY condition thus leads to
\begin{equation}
\mathcal{A}+\mathcal{B}\log(-i\xi)+\mathcal{C}\frac{\xibar^{1/3}}{\xi^{2/3}}=0,\label{NCeqnWarped}
\end{equation}
 with $\mathcal{A}$ and $\mathcal{B}$ as before and
\begin{equation}
\mathcal{C}=3\, C_{w}(F-\tau H)\cdot a_{0}.
\end{equation}

From this, we can see the rough influence of warping on the distribution
of vacua. In the unwarped case, we expect to find vacua $10^{-100}$
or so away from the conifold with fluxes yielding $|\mathcal{A}|\sim100|\mathcal{B}|$
(which with fluxes constrained to lie in $(0,100$ is about the maximum
order of magnitude difference that we expect.) If however, $C_{w}\sim10^{-20}$,
then for $|\xi|\sim10^{-100}$, the warp term contribution is on the
order of $10^{10}$, swamping the logarithmic contribution and requiring
fluxes $|\mathcal{A}|\sim10^{10}$ which lies beyond the range we
consider.

In the region of strong warping where the logarithmic term is dominated
by the warping term, the distance of a vacuum from the conifold point
is thus set by $|\mathcal{C}/\mathcal{A}|^{3}$. Given that $\mathcal{A}$
is at maximum of roughly 100 or so, the constant $C_{w}$, and thus,
the overall volume of the Calabi-Yau, determines how near the conifold
vacua lie. This can dramatically truncate the range---since the assumption
of large but finite volume is well satisfied by volumes of order $10^{20}$,
but in those cases, vacua will not show up much closer than $10^{-60}$.
We can get vacua at around $10^{-120}$ by taking a volume of order
$10^{40}$, but in the absence of warping, vacua as far in as $10^{-200}$
are expected. Thus, warping pushes vacua away from the conifold point.

\subsection{Monte-Carlo vacua}

For the numerical analysis, we use the Calabi-Yau manifold labeled
model 3 in the appendix of \cite{Conifunneling}. This family of Calabi-Yau
can be expressed as a locus of octic polynomials in $\mathbb{WP}^{4,1,1,1,1}$.
The corresponding orientifold arises from a certain limit of F-theory
compactified on a Calabi-Yau fourfold hypersurface in $\mathbb{WP}^{12,8,1,1,1,1},$
following the methods of \cite{SenFTheory}, and briefly described
in \cite{KachruCY3Fold}. For our purposes, we use the fact that the
fourfold has Euler characteristic $\chi=23328$, which implies that
$L_{\textrm{max}}=\chi/24=972$ for the tadpole condition for flux
compactification on the corresponding orientifolded 3-fold.

Since the warped form of the near conifold equation is not as simple
to solve as in the unwarped case, a slightly more involved approach
is necessary. We begin by defining two real variables $\rho$ and
$\theta$ such that
\begin{equation}
-i\xi=\rho^{3}e^{i\theta}
\end{equation}

We take $\rho\geq0$ and $0\leq\theta\leq2\pi$. In terms of these
variables, eqn (\ref{NCeqnWarped}) and its complex conjugate expression
take the form
\begin{eqnarray}
\mathcal{A}+3\mathcal{B}\ln(\rho)+i\mathcal{B}\theta+\frac{\mathcal{C}}{\rho}e^{-i\theta} & = & 0\\
\overline{\mathcal{A}}+3\overline{\mathcal{B}}\ln(\rho)-i\overline{\mathcal{B}}\theta+\frac{\overline{\mathcal{C}}}{\rho}e^{i\theta} & = & 0
\end{eqnarray}

Multiplying the first equation by $\overline{\mathcal{B}}$ and the
second one by $\mathcal{B}$, and then adding and subtracting the
two, we find two purely real or imaginary equations. Letting $\mathcal{A}=ae^{i\alpha}$,
$\mathcal{B}=be^{i\beta}$, and $\mathcal{C}=ce^{i\gamma}$, we have
\begin{eqnarray}
a\sin(\alpha-\beta)+b\theta+\frac{c}{\rho}\sin(\gamma-\beta-\theta) & = & 0\label{first}\\
a\cos(\alpha-\beta)+3b\log(\rho)+\frac{c}{\rho}\cos(\gamma-\beta-\theta) & = & 0\label{second}
\end{eqnarray}

We now solve for $\rho$ in terms of $\theta$ and then numerically
solve the final equation for $\theta$. It seems natural to solve
equation (\ref{first}) for $\rho$ since it is a linear equation.
However, this approach fails in the limit $C_{w}\rightarrow0$ since
then $c\rightarrow0$ too. Instead we solve for $\rho$ in equation
(\ref{second}). One can rearrange the equation as
\begin{equation}
\rho e^{\Gamma}\log(\rho e^{\Gamma})=-\frac{ce^{\Gamma}}{3b}\cos(\gamma-\beta-\theta)
\end{equation}

Here we have defined the constant $\Gamma=\frac{a\cos(\alpha-\beta)}{3b}$.
This is of the form $x\log(x)=y$ which has the solution $x=y/W(y)$
where $W(y)$ is the Lambert $W$-function. We therefore find
\begin{equation}
\rho(\theta)=\frac{-c\cos(\gamma-\beta-\theta)}{3bW(-\frac{ce^{\Gamma}}{3b}\cos(\gamma-\beta-\theta))}\label{rho}
\end{equation}

Consider equation (\ref{first}), which now only depends on $\theta$.
Under the assumption that there is only one near conifold vacuum for
each set of fluxes, the left hand side must either start out positive,
and go negative or vice versa. To find the zero-crossing, we divide
the region $[0,2\pi]$ into two equally pieces and then determine
in which region (if any) equation (\ref{first}) changes sign. If
such a region is found, we apply the same method to that region, splitting
it into two smaller intervals, continuing in this way until we reach
a predetermined level of accuracy. There are two relevant comments.
First, in equation (\ref{rho}) it is not clear that the value of
$\rho$ is real, or even positive. We must therefore exclude the regions
where $\rho$ is either negative or complex. Fortunately, if $\rho$
is real, it is never negative since $W(x)$ must have the same sign
as $x$. A necessary and sufficient condition for $\rho$ to be real
is that the argument of the Lambert $W$ function is greater than
or equal to $-1/e$. This means that the relevant region to begin
with may not be the entire interval $[0,2\pi]$. Second, it turns
out that the Lambert $W$ function has two real branches for arguments
between $-1/e$ and $0$. Thus, both of these branches must be considered.

\begin{figure}
\includegraphics[width=16cm]{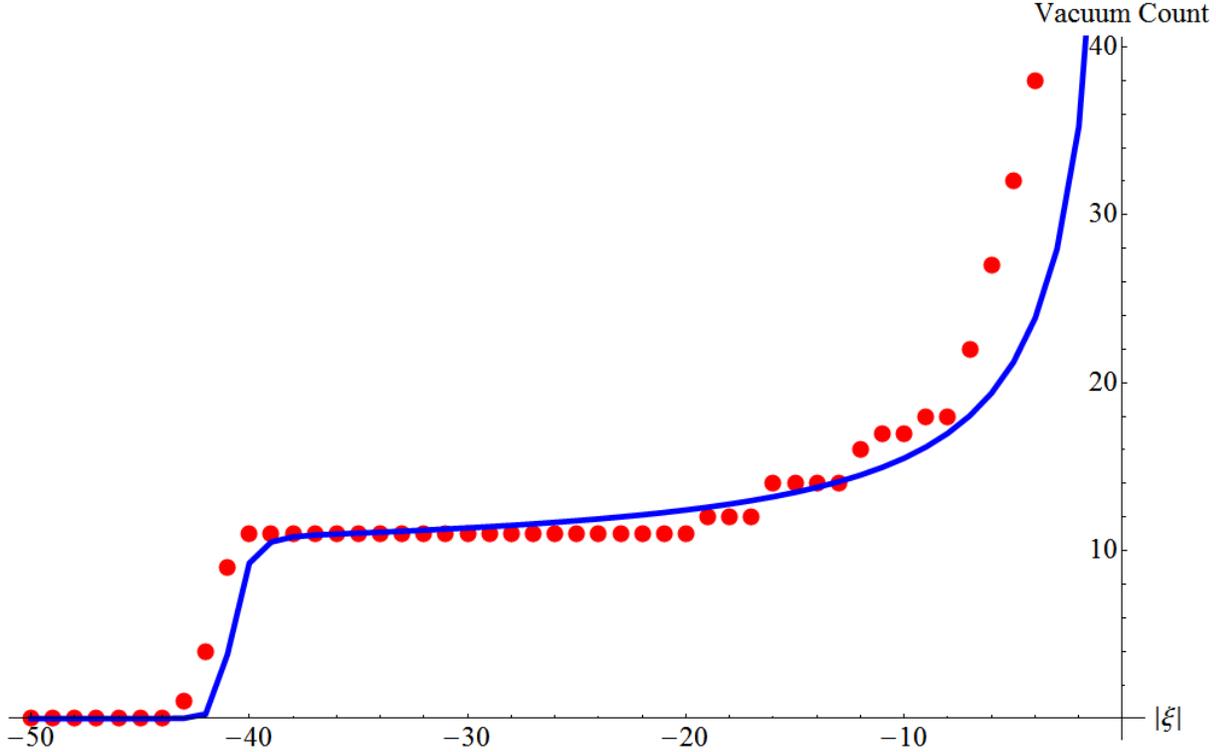}

\caption{\label{fig:NumericsVsTheory}A comparison between numerical and analytical
distributions. Red circles mark the numerical data while the blue
curve is the integrated analytical distribution. Distance from the
conifold $\left|\xi\right|$ is plotted on a log scale on the horizontal
axis, while the vacuum count is plotted on the vertical axis.}
\end{figure}

To better compare the numerical and analytical and numerical distributions,
we fix $\tau$ and then select a random sets of fluxes $F$ and $H$
consistent with our choice of $\tau$ and satisfying the tadpole condition,
$F\cdot Q\cdot H\leq L_{max}$. For the particular model we consider,
$L_{max}=972$, and we display a run using $\tau=2i$, and $C_{w}=10^{-15}$
in figure \ref{fig:NumericsVsTheory}. The figure shows a numerical
run compared to the analytical distribution. We plot the vacuum count
and integrated analytical distribution as measured around the conifold
point using a log scale for the distance from the conifold. As is
evident from the figure, the count receives two major contributions:
the one farther away from the conifold point is the usual contribution
that is present without warping. However, we also see a major contribution
much closer to the conifold at a distance roughly on the order of
$C_{w}^{3}$. This contribution is due to the strong warping effects
and is matched by the cumulative analytical results.

\section{Discussion}

We've analyzed the distribution of flux vacua in the vicinity of the
conifold point, including the effects of warping, and confirmed our
results by a direct numerical Monte Carlo search. In comparison with
the well known results, that don't include warping, we find a significant
dilution of vacua in close proximity to the conifold, with the proximity
scale set by the volume of the Calabi-Yau compactification.

One complication in the analytical approach, relative to the unwarped
case, is the need to bound the fluxes -- a physically sensible requirement
but one that can be avoided in the unwrapped analysis, yielding the
geometrical result of \cite{DouglasAshok,DouglasDenef}. It would
be interesting to see whether the warped distribution of vacua can
once again be related to intrinsic properties of the moduli space
through a more complete geometrical treatment, likely requiringing
careful consideration of the generalized complex geometry of conformal
Calabi-Yau spaces \cite{Koerber,Martucci}.

\subsection*{Acknowledgements}

We thank Michael Douglas, Saswat Sarangi, Gary Shiu, and I-Sheng Yang
for helpful discussions. Pontus Ahlqvist is partly supported by a
graduate fellowship from the Sweden America Foundation. This work
is supported in part by DOE grant DE-FG02-92ER40699, FQXi grant RFP1-06-19,
and STARS grant CHAPU G2009-30 7557.

\appendix

\section{Appendix}

\subsection{Covariant Derivatives\label{sub:Covariant-Derivatives}}

We start with the standard definition of the Kähler potential
\[
e^{-K}=\int\widehat{\Omega}_{4}\wedge\overline{\widehat{\Omega}}_{4}
\]
 A rescaling of the holomorphic 4-form $\widehat{\Omega}_{4}\rightarrow e^{f(z)}\widehat{\Omega}_{4}$
implies that $K\rightarrow K-f(z)-\overline{f(z)}$. The covariant
derivative is defined so that it is covariant under such rescalings:
\[
D_{a}\widehat{\Omega}_{4}\rightarrow e^{f(z)}D_{a}\widehat{\Omega}_{4}=\left(D_{a}-\partial_{a}f\right)\left(e^{f(z)}\widehat{\Omega}_{4}\right)
\]
 implying that $D_{a}\widehat{\Omega}_{4}=\left(\partial_{a}+K_{a}\right)\widehat{\Omega}_{4}$.
Note also that the holomorphic covariant derivative annihilates antiholomorphic
objects, i.e. $D_{a}\overline{\widehat{\Omega}}_{4}=\partial_{a}\overline{\widehat{\Omega}}_{4}=0$.
We see then that
\[
D_{a}e^{-K}=\left(\partial_{a}+K_{a}\right)e^{-K}=0
\]
 Notice that covariance dictates that
\[
D_{a}e^{K}=\left(\partial_{a}-K_{a}\right)e^{K}=0
\]

In addition to this Kähler scaling structure, the complex structure
moduli space is a manifold with a natural Kähler metric $K_{a\bar{b}}=\partial_{a}\partial_{\bar{b}}K$.
We can thus define a metric compatible connection via
\[
\Gamma_{bc}^{a}=K^{a\bar{d}}K_{\bar{d}bc}
\]
 where $K_{\bar{d}bc}=\partial_{c}K_{\bar{d}b}$. Note that the connection
components with mixed holomorphic and antiholomorphic indices vanish.
By suitably extending the covariant derivative $D_{a}$, we can ensure
that it transforms covariantly under both Kähler rescalings and coordinate
transformations on the complex moduli space. In particular, since
the connection is metric compatible, $D_{a}K_{b\bar{c}}=0$.

The superpotential
\[
\widehat{W}=\int G_{4}\wedge\widehat{\Omega}_{4}
\]
 scales as $\widehat{\Omega}_{4}$, and thus
\[
D_{a}\widehat{W}=\partial_{a}\widehat{W}+K_{a}\widehat{W}
\]
 A supersymmetric vacuum satisfies the conditions $D_{a}\widehat{W}=0$.
The components of the fermion mass matrix are
\[
\partial_{a}D_{b}\widehat{W}=\partial_{a}\partial_{b}\widehat{W}+K_{ab}\widehat{W}+K_{b}\widehat{W}_{a}
\]
 Notice that at a generic point in the moduli space the quantity
\[
D_{a}D_{b}\widehat{W}=\partial_{a}D_{b}\widehat{W}+K_{a}D_{b}\widehat{W}-\Gamma_{ab}^{c}D_{c}\widehat{W}=\partial_{a}D_{b}\widehat{W}+K_{a}D_{b}\widehat{W}-K_{ab\bar{d}}K^{\bar{d}c}D_{c}\widehat{W}
\]
 does not equate to the fermion mass matrix. However, the extra terms
drop out at supersymmetric vacua.

Now let $\widehat{\Omega}_{4}\rightarrow\Omega_{4}=e^{K/2}\widehat{\Omega}_{4}$
and similarly for the (0,4)-form. Notice that the scaling properties
of $\Omega_{4}$ imply that
\[
D_{a}\Omega_{4}=\left(\partial_{a}+\frac{1}{2}K_{a}\right)\Omega_{4}=\left(\partial_{a}+\frac{1}{2}K_{a}\right)\left(e^{K/2}\widehat{\Omega}_{4}\right)=e^{K/2}\left(\partial_{a}+K_{a}\right)\widehat{\Omega}_{4}=e^{K/2}D_{a}\widehat{\Omega}_{4}
\]
 The rescaled (4,0) and (0,4) forms are convenient since they remove
factors of $e^{K}$ from various expressions. In particular we have
\[
\int\Omega_{4}\wedge\overline{\Omega}_{4}=1
\]

We can also go to an orthonormal frame by introducing vielbeins $\delta_{A\overline{B}}=e_{A}^{a}e_{\overline{B}}^{\overline{b}}K_{a\bar{b}}$.
The covariant derivative must be extended so as to keep the vielbeins
covariantly constant:
\[
D_{a}e_{B}^{b}=\partial_{a}e_{B}^{b}+K^{b\bar{d}}K_{\bar{d}ac}e_{B}^{c}-\omega_{aB}{}^{C}e_{C}^{b}=0
\]
 implying that
\[
\omega_{AB}{}^{C}=e_{A}^{a}\omega_{aB}{}^{C}=e_{b}^{C}e_{A}^{a}\partial_{a}e_{B}^{b}+K^{b\bar{d}}K_{\bar{d}ac}e_{A}^{a}e_{B}^{c}e_{b}^{C}
\]
 Given these definitions, we can now go to rescaled expressions in
the orthonormal frame:
\[
D_{A}D_{B}W=\partial_{A}D_{B}W+K_{A}D_{B}W-\omega_{AB}{}^{C}D_{C}W
\]
 once again, the expression above agrees with the fermion mass matrix
components evaluated at a vacuum.

\subsection{Computations}

In this section, we provide derivations for the equations (\ref{eq:Xdef})-(\ref{eq:D0bDI}).
To do so, recall from equation (\ref{eq:G4OmegaBasis}) that the flux
four form written in the basis $\mathcal{B}=\{\Omega_{4},D_{A}\Omega_{4},D_{\underline{0}}D_{I}\Omega_{4},\overline{\Omega}_{4},\overline{D_{A}\Omega}_{4},\overline{D_{\underline{0}}D_{I}\Omega}_{4}\}$
is
\begin{equation}
G_{4}=\overline{X}\Omega_{4}-\overline{Y}{}^{A}D_{A}\Omega_{4}+\overline{Z}{}^{I}D_{\underline{0}}D_{I}\Omega_{4}+\text{c.c.}
\end{equation}

It's useful to note that $D_{A}D_{B}\Omega_{4}$ is a (2,2)-form.
In fact, given the nature of our Calabi-Yau 4-fold, essentially factorizing
into a 3-fold and a torus, this (2,2)-form can be decomposed as $(2,2)=(1,0)\wedge(1,2)\oplus(0,1)\wedge(2,1)$.
To see this, note that $D_{A}\Omega_{4}$ could be a mixture of a
$(3,1)\oplus(4,0)$, but the (4,0) component vanishes:
\[
\int_{\mathcal{M}}D_{A}\Omega_{4}\wedge\overline{\Omega}_{4}=D_{A}\left(\int_{\mathcal{M}}\Omega_{4}\wedge\overline{\Omega}_{4}\right)-\int_{\mathcal{M}}\Omega_{4}\wedge D_{A}\overline{\Omega}_{4}=0
\]
 where the two terms vanish given the properties of the covariant
derivative defined above. Similarly $D_{A}D_{B}\Omega_{4}$ could
in principle have $(2,2)\oplus(3,1)\oplus(4,0)$ structure. However,
\[
\int_{\mathcal{M}}D_{A}D_{B}\Omega_{4}\wedge\overline{\Omega}_{4}=D_{A}\left(\int_{\mathcal{M}}D_{B}\Omega_{4}\wedge\overline{\Omega}_{4}\right)-\int_{\mathcal{M}}D_{B}\Omega_{4}\wedge D_{A}\overline{\Omega}_{4}=0
\]
 implying that there is no $(4,0)$ component. Furthermore
\[
\int_{\mathcal{M}}D_{A}D_{B}\Omega_{4}\wedge\overline{D}_{\overline{C}}\overline{\Omega}_{4}=D_{A}\left(\int_{\mathcal{M}}D_{B}\Omega_{4}\wedge\overline{D}_{\overline{C}}\overline{\Omega}_{4}\right)-\int_{\mathcal{M}}D_{B}\Omega_{4}\wedge D_{A}\overline{D}_{\overline{C}}\overline{\Omega}_{4}
\]
 the first term on the left-hand-side is equal to $D_{A}\delta_{B\overline{C}}=0$.
The second term becomes
\[
\delta_{A\overline{C}}\int_{\mathcal{M}}D_{B}\Omega_{4}\wedge\overline{\Omega}_{4}=0
\]
 which shows that there is no $(3,1)$ component in $D_{A}D_{B}\Omega_{4}$.
We now turn to the identities of interest.\\

\begin{itemize}
\item $W=X$ \\

By the definition of the superpotential, we have
\begin{eqnarray}
W & = & \int_{\mathcal{M}}G_{4}\wedge\Omega_{4}\nonumber \\
 & = & \int_{\mathcal{M}}\left(\overline{X}\Omega_{4}-\overline{Y}{}^{A}D_{A}\Omega_{4}+\overline{Z}{}^{I}D_{\underline{0}}D_{I}\Omega_{4}+\text{c.c.}\right)\wedge\Omega_{4}=X
\end{eqnarray}

In the last step we used the orthonormality of the basis.

\item $D_{A}W=Y_{A}$ \\

Once again, we will use the orthonormality of the basis. In particular
we have (since $G_{4}$ is independent of the moduli)
\begin{eqnarray}
D_{A}W & = & \int_{\mathcal{M}}G_{4}\wedge D_{A}\Omega_{4}\nonumber \\
 & = & \int_{\mathcal{M}}\left(\overline{X}\Omega_{4}-\overline{Y}{}^{A}D_{A}\Omega_{4}+\overline{Z}{}^{I}D_{\underline{0}}D_{I}\Omega_{4}+\text{c.c.}\right)\wedge D_{A}\Omega_{4}\nonumber \\
 & = & -Y^{\bar{B}}\int_{\mathcal{M}}\overline{D}_{\bar{B}}\overline{\Omega}_{4}\wedge D_{A}\Omega_{4}=+Y_{A}
\end{eqnarray}
 In the last step we again used ($\int_{\mathcal{M}}\overline{D}_{\bar{B}}\overline{\Omega}_{4}\wedge D_{A}\Omega_{4}=-\delta_{\bar{B}A}$).

\item $D_{\underline{0}}D_{\underline{0}}W=0$

We have
\begin{eqnarray}
D_{\underline{0}}D_{\underline{0}}W & = & \int_{\mathcal{M}}G_{4}\wedge D_{\underline{0}}D_{\underline{0}}\Omega_{4}
\end{eqnarray}
 Now $D_{\underline{0}}\Omega_{4}$ is a $(0,1)\wedge(3,0)$-form.
In fact, we see that
\[
D_{\tau}\widehat{\Omega}_{1}=\left(\partial_{\tau}+K_{\tau}\right)\left(\alpha-\tau\beta\right)=K_{\tau}\left(\alpha-\overline{\tau}\beta\right)=K_{\tau}\overline{\widehat{\Omega}}_{1}
\]
 where we have used $K_{\tau}=-1/\left(\tau-\overline{\tau}\right)$.
Using the fact that the vielbein $e_{\underline{0}}^{0}=1/K_{\tau}$,
we have $D_{\underline{0}}\Omega_{4}=\overline{\Omega}_{1}\wedge\Omega_{3}$,
however we know that $D_{\underline{0}}\overline{\Omega}_{1}=0$,
so the identity holds.

\item $D_{\underline{0}}D_{I}W=Z_{I}$

This identity follows from orthonormality:
\begin{equation}
D_{\underline{0}}D_{I}W=\int_{\mathcal{M}}G_{4}\wedge D_{\underline{0}}D_{I}\Omega=\int_{\mathcal{M}}\left(\overline{X}\Omega_{4}-\overline{Y}{}^{A}D_{A}\Omega_{4}+\overline{Z}{}^{I}D_{\underline{0}}D_{I}\Omega_{4}+\text{c.c.}\right)\wedge D_{\underline{0}}D_{I}\Omega=Z_{I}
\end{equation}

\item $D_{I}D_{J}W=\mathcal{F}_{IJK}\overline{Z}^{K}$

We will again use the definition for $G_{4}$
\begin{equation}
D_{I}D_{J}W=\int_{\mathcal{M}}\left(\overline{X}\Omega_{4}-\overline{Y}{}^{A}D_{A}\Omega_{4}+\overline{Z}{}^{I}D_{\underline{0}}D_{I}\Omega_{4}+\text{c.c.}\right)\wedge D_{I}D_{J}\Omega_{4}
\end{equation}

As discussed above $D_{A}D_{B}\Omega_{4}$ is a (2,2)-form which breaks
up as $(1,0)\wedge(1,2)\oplus(0,1)\wedge(2,1)$. Now, $D_{I}D_{J}\Omega_{4}$
is precisely a $(1,0)\wedge(1,2)$-form, so the covariant derivatives
only act on the 3-fold factor. The only pieces of the integral above
that can yield a non-zero result must be of the form $(0,1)\wedge(2,1)$,
which are thus proportional to $D_{\underline{0}}D_{I}\Omega_{4}$.
This leaves us with
\[
D_{I}D_{J}W=\overline{Z}^{K}\int_{\mathcal{M}}D_{\underline{0}}D_{K}\Omega_{4}\wedge D_{I}D_{J}\Omega_{4}
\]
 The $D_{K}$ and $D_{\underline{0}}$ derivatives commute, so we
have
\[
D_{I}D_{J}W=\overline{Z}^{K}D_{K}\left(\int_{\mathcal{M}}D_{\underline{0}}\Omega_{4}\wedge D_{I}D_{J}\Omega_{4}\right)-\overline{Z}^{K}\int_{\mathcal{M}}D_{\underline{0}}\Omega_{4}\wedge D_{K}D_{I}D_{J}\Omega_{4}
\]
 The first term on the right-hand-side vanishes due to orthonormality
since $D_{\underbar{0}}\Omega_{4}$ is a (3,1)-form while $D_{I}D_{J}\Omega_{4}$
is a $(1,0)\wedge(1,2)$-form. Factorizing $\Omega_{4}=\Omega_{1}\wedge\Omega_{3}$,
we see that $D_{\underline{0}}\Omega_{4}=\overline{\Omega}_{1}\wedge\Omega_{3}$.
The integral over the torus will simply yield a factor of $-i$, leaving
us with
\[
D_{I}D_{J}W=i\overline{Z}^{K}\int_{\mathcal{CY}}\Omega_{3}\wedge D_{K}D_{I}D_{J}\Omega_{3}
\]
 However, pulling out all scaling factors and vielbeins, we see that
the resulting derivatives can all be converted to partials. This allows
us to rearrange the ordering and gives
\[
D_{I}D_{J}W=i\overline{Z}^{K}\int_{\mathcal{CY}}\Omega_{3}\wedge D_{I}D_{J}D_{K}\Omega_{3}=\mathcal{F}_{IJK}\overline{Z}^{K}
\]

\item $\bar{D}_{\bar{A}}D_{B}W=\delta_{\bar{A}B}X$

First consider
\begin{eqnarray*}
\bar{D}_{\bar{a}}D_{b}\widehat{W} & = & \bar{\partial}_{\bar{a}}(\partial_{b}+K_{b})\widehat{W}=(\bar{\partial}_{\bar{a}}\partial_{b}+K_{b\bar{a}}+K_{b}\bar{\partial}_{\bar{a}})\widehat{W}=K_{\overline{a}b}\widehat{W}
\end{eqnarray*}

The first and last term vanish since $\widehat{W}$ is holomorphic
in the moduli. Then, since $W=X$, by reintroducing the scaling factor
$e^{K/2}$ and the vielbeins, we have
\begin{equation}
\bar{D}_{\bar{A}}D_{B}W=\delta_{\bar{A}B}X
\end{equation}

\item $\bar{D}_{\bar{\underline{0}}}D_{I}W=0$

We can easily see this by noting that the outer derivative is a regular
partial derivative and that this commutes with the inner derivative.
Then, since $\hat{W}$ is holomorphic in $\tau$, $\bar{\partial}_{\bar{0}}$
sends the expression to zero.

\end{itemize}

\end{document}